\documentclass[
	 reprint,
	 longbibliography,
	 amsmath,
	 amssymb,
	 aps,
	 pre,
]{revtex4-1}

\usepackage{dcolumn}% Align table columns on decimal point
\usepackage{bm}% bold math
\usepackage{amssymb}
\usepackage{amsmath}
\usepackage{amsfonts}
\usepackage{cases}
\usepackage{graphicx,color}
\usepackage{multirow}
\usepackage{makecell}
\usepackage{mathtools}
\usepackage[justification=raggedright]{caption}

\newcolumntype{R}[1]{>{\raggedleft\arraybackslash}p{#1}}

\definecolor{r}{rgb}{1,0,0}   
\definecolor{g}{rgb}{0,1,0}   
\definecolor{b}{rgb}{0,0,1}
\definecolor{purple}{rgb}{0.808,0.454,0.718}
\begin{document}

% Use the \preprint command to place your local institutional report
% number in the upper righthand corner of the title page in preprint mode.
% Multiple \preprint commands are allowed.
% Use the 'preprintnumbers' class option to override journal defaults
% to display numbers if necessary
%\preprint{}

%Title of paper
\title{The Spectrum of Structure for Jammed and Unjammed Soft Disks}

% repeat the \author .. \affiliation  etc. as needed
% \email, \thanks, \homepage, \altaffiliation all apply to the current
% author. Explanatory text should go in the []'s, actual e-mail
% address or URL should go in the {}'s for \email and \homepage.
% Please use the appropriate macro for each each type of information

% \affiliation command applies to all authors since the last
% \affiliation command. The \affiliation command should follow the
% other information
% \affiliation can be followed by \email, \homepage, \thanks as well.
\author{A. T. Chieco$^1$, M. Zu$^2$, A. J. Liu$^1$, N. Xu$^2$, D. J. Durian$^1$}
%\email[]{Your e-mail address}
%\homepage[]{Your web page}
%\thanks{}
%\altaffiliation{}
\affiliation{
   $^1$Department of Physics and Astronomy, University of Pennsylvania, Philadelphia, PA 19104-6396, USA  \\
   $^2$Department of Physics, University of Science and Technology of China, Hefei 230026, People's Republic of China
}

%Collaboration name if desired (requires use of superscript address
%option in \documentclass). \noaffiliation is required (may also be
%used with the \author command).
%\collaboration can be followed by \email, \homepage, \thanks as well.
%\collaboration{}
%\noaffiliation

\date{\today}%~~***DRAFT***}

\begin{abstract}
We investigate the short, medium, and long-range structure of soft disk configurations for a wide range of area fractions and simulation protocols by converting the real-space spectrum of volume fraction fluctuations for windows of width $L$ to the distance $h(L)$ from the window boundary over which fluctuations occur.  Rapidly quenched unjammed configurations exhibit size-dependent super-Poissonian long-range features that, surprisingly, approach the totally-random limit even close to jamming.  Above and just below jamming, the spectra exhibit a plateau, $h(L)=h_e$, for $L$ larger than particle size and smaller than a cutoff $L_c$ beyond which there are long-range fluctuations.  The value of $h_e$ is independent of protocol and characterizes the putative hyperuniform limit.  This behavior is compared with that for Einstein solids, with and without hyperuniformity-destroying defects.  We find that key structural features of the particle configurations are more evident, as well as easier and more intuitive to quantify, using the real-space spectrum of hyperuniformity lengths rather than the spectral density.
\end{abstract}

% insert suggested PACS numbers in braces on next line
%45.70.-n Granular systems
%47.57.Gc Granular flow
%83.80.Fg Granular solids
%81.70.Bt Mechanical testing, impact tests, static and dynamic loads

% insert suggested keywords - APS authors don't need to do this
%\keywords{}

%\maketitle must follow title, authors, abstract, \pacs, and \keywords
\maketitle

% body of paper here - Use proper section commands
% References should be done using the \cite, \ref, and \label commands
%\section{}
% Put \label in argument of \section for cross-referencing
%\section{\label{}}
%\subsection{}
%\subsubsection{}

% If in two-column mode, this environment will change to single-column
% format so that long equations can be displayed. Use
% sparingly.
%\begin{widetext}
% put long equation here
%\end{widetext}

%--------------------------------------------------------------------------------------------------------------------------------------------------------------------------------------------

\section{Introduction}

Systems with strongly suppressed long-range density fluctuations are said to be {\it hyperuniform} \cite{TorquatoPRE2003, ZacharyJSM2009}.  This can indicate hidden order, giving dramatic properties such as isotropic optical band gaps \cite{FlorescuPNAS2009, ManPNAS2013, MullerAOM2014, Scheffold2016} and criticality in dynamical absorbing state transitions~\cite{HexnerPRL2015, TjhunPRL2015, WeijsPRL2015}. It is conjectured that all strictly jammed saturated hard-particle packings are hyperuniform, no matter what the particles' sizes or shapes \cite{TorquatoPRE2003, DonevPRL2005, TorquatoStillingerRMP10, ZacharyPRL2011, DreyfusPRE2015, AtkinsonPRE2016}.  However, some simulations indicate otherwise \cite{IkedaPRE2015, WuPRE2015, ParisiPRE2017, Ozawa2017}.  This suggests long-ranged structural features not evident in measures of the local packing environment (e.g.~\cite{ARCMP, CorwinPRL14, SchoderTurkEL15, CubukPRL15, RieserPRL16}).  We aim to address the nature and consequences of such nonlocal structure, at various length scales, as an important issue beyond the yes/no question of whether or a system is hyperuniform.

The spectrum of structural features can, in principle, be probed using the same tools used to diagnose hyperuniformity.  These tools include the spectral density $\chi(q)$, equal to the structure factor for monodisperse systems, and the variance $\sigma_\phi^2(L)$ in the set of local volume fractions measured in randomly-placed windows of width $L$.  At long lengths, scaling of $\chi(q)\sim q^\epsilon$ with $\epsilon \le 1$ corresponds to $\sigma_\phi^2(L)\sim 1/L^{d+\epsilon}$ where $d$ is dimensionality; for $\epsilon \ge 1$, $\chi(q)\sim q^\epsilon$ corresponds to $\sigma_\phi^2(L)\sim 1/L^{d+1}$.  Ordinary systems exhibit Poissonian fluctuations with $\epsilon=0$.  By contrast, hyperuniform ones have $\chi(0^+)=0$ and $\epsilon>0$; for $\epsilon \ge 1$ we say the system is strongly hyperuniform.  Until now, interest has been solely on asymptotics such as fits for $\epsilon$ and $\chi(0^+)$, which tell us only whether or not a system is hyperuniform.  Here we shift focus onto the {\it values} of $\chi(q)$ and $\sigma_\phi^2(L)$ at each length. In particular, we show that $\sigma_\phi^2(L)$ can be converted easily into a ``hyperuniformity disorder length," $h(L)$  \cite{ATCpixel, DJDhudls}, and that $h(L)$ can be interpreted straightforwardly to give novel insight into structure not only at the longest length scales but also at short and intermediate scales.

%------------------------------------------------------------------- DEFINITIONS, SCALING, AND BOUNDS:

\section{Definitions and Bounds}

To extract meaning from values, we first normalize $\chi(q)$ and $\sigma_\phi^2(L)$ in $d$ dimensions relative to a  ``Poisson pattern" where particles are placed totally at random.  Throughout we use a central-point representation in which the volume $v_j$ of particle $j$ is encoded at the location ${\bf r}_j$ of its center \cite{extendedparticle}.  For the spectral density, a suitable definition is 
\begin{equation}
	\chi(q) \equiv {  \left(  \sum v_j e^{i {\bf q}\cdot{\bf r}_j}~\sum v_k e^{-i {\bf q}\cdot{\bf r}_k}\right) }/{\sum v_j^2 }
\label{chidef}
\end{equation}
where $q=|{\bf q}|$ for isotropic packings and the sums are over all particles in the system.  With this normalization, which has been neglected in the past, Poisson patterns have $\chi(q)=1$; this is important because some insight into structure at a given $q$ can then be extracted from how far $\chi(q)$ lies below the nominal upper bound of 1.   

For the volume fraction variance, $\sigma_\phi^2(L)$ is calculated from a set of local volume fractions $\sum v_i N_i/V_\Omega$, where $N_i$ is the number of particles of species $i$ whose centers lie inside a particular placement of a window of volume $V_\Omega\sim L^d$ and the sum is over species. This is the real space definition for the central point representation. For a Poisson pattern, we compute $\sigma_{N_i}^2 =\overline{N_i} = (\phi_i / v_i)V_\Omega$, where $\phi_i$ is the true volume fraction of species $i$ in the entire sample; this gives
\begin{equation}
	\frac{\sigma_\phi^2(L)}{\phi} = \frac{\langle v \rangle }{V_\Omega}
\label{poisson}
\end{equation}
as a nominal upper bound.  Here, $\langle v \rangle =\sum \phi_i v_i/\phi$ is the $\phi_i$-weighted average particle volume and $\phi=\sum\phi_i$ is the true volume fraction occupied by all particles.   Also, for repulsive particles or finite systems, a minimum center-to-center distance exists and hence small enough windows contain either zero or one particle center.  This leads to
\begin{equation}
	\frac{\sigma_\phi^2(L)}{\phi} = \frac{\langle v \rangle }{V_\Omega} - \phi
\label{separated}
\end{equation}
as the ``separated-particle" lower bound \cite{ATCpixel, DJDhudls}.

\begin{figure}[ht]
\includegraphics[width=3in]{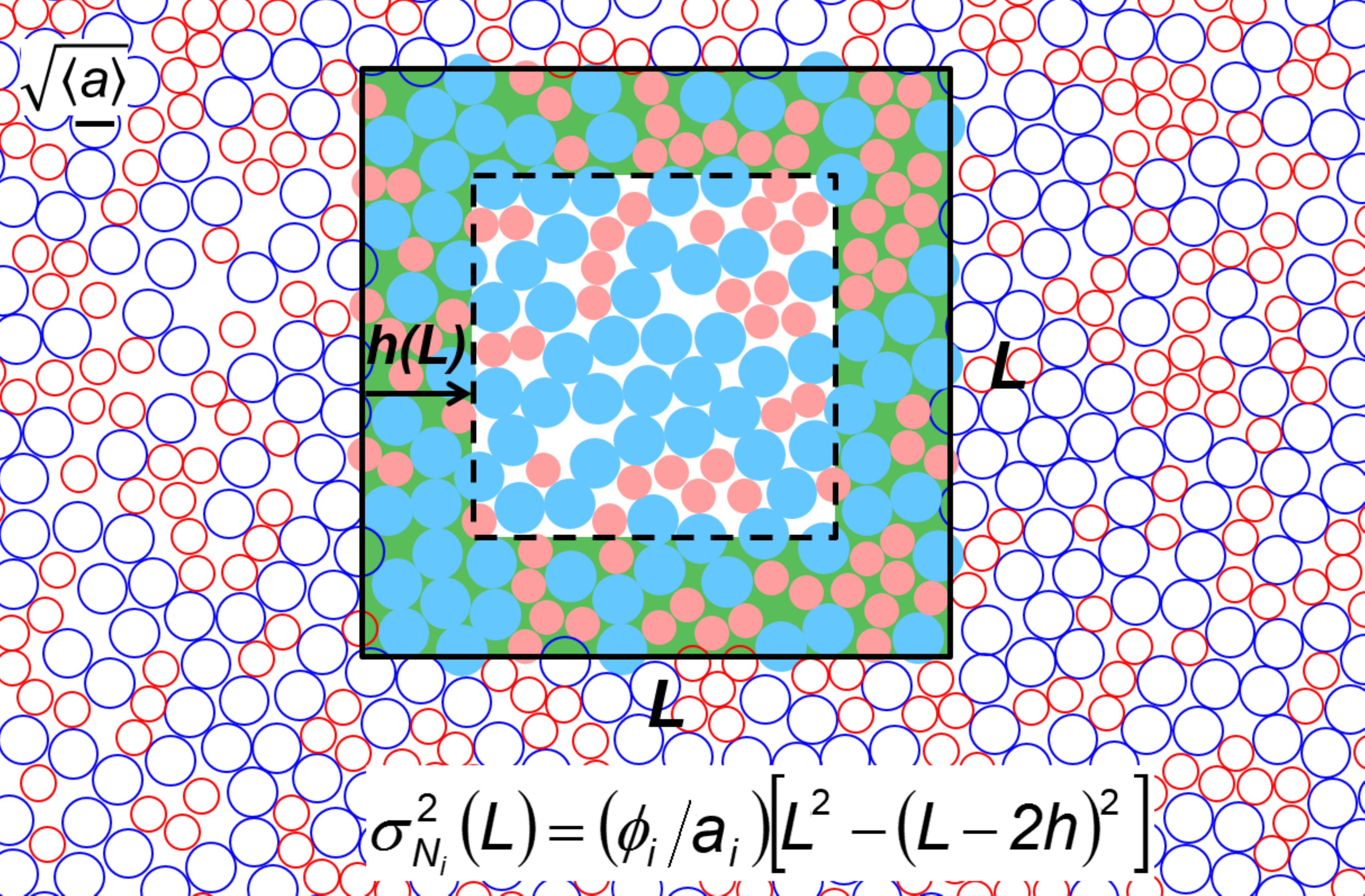}
\caption{Unjammed bidisperse disks at area fraction $\phi=0.70$ made by rapid quench.  The area fraction variance is controlled by the number of particles in the shaded region of thickness $h(L)$, averaged over window placements.  As depicted here for $L=15 \sqrt{\left<a\right>}$, the hyperuniformity disorder length is $h(L)=3 \sqrt{\left<a\right>}$ where $\langle a\rangle$ is the area-fraction weighted average disk area, equal to 0.81 large disk diameters.}
\label{hdefpic}
\end{figure}

To supplement these bounds we seek a measure of order that is independent of $L$ if the system is hyperuniform, with fluctuations understood as due to particles at the surface of the measuring windows \cite{TorquatoPRE2003}.  Since particle centers do not actually lie {\it on} the window surface, it is more appropriate to picture fluctuations as determined by the average number of particles whose centers lie {\it within some distance} $h$ of the surface.  Taking the windows to be hypercubic with volume $V_\Omega=L^d$, as depicted for $d=2$ in Fig.~\ref{hdefpic}, and inserting $\sigma_{N_i}^2 = (\phi_i / v_i)[L^d - (L-2h)^d]$ into $\sigma_\phi^2=\sum \sigma_{N_i}^2 v_i^2/V_\Omega^2$, leads to the following definition of $h(L)$ in terms of the measured variance:
\begin{eqnarray}
	\frac{ \sigma_\phi^2(L) }{\phi} &\equiv& \frac{\langle v\rangle}{L^d}\left\{ 1 - \left[1-\frac{h(L)}{L/2} \right]^d\right\} \label{hdef}, \\
						     &\approx& 2d \frac{ \langle v\rangle h(L) }{ L^{d+1} }\ {\rm for}\ L\gg h(L). \label{hdefapprox}
\end{eqnarray}
Accordingly, smaller $h(L)$ means more uniformity, larger $h(L)$ means more disorder, and $\sigma_\phi^2(L)\sim1/L^{d+\epsilon}$ corresponds to $h(L)\sim L^{1-\epsilon}$.  Poissonian fluctuations ($\epsilon=0$) correspond to $h(L) \sim L$; the upper bound is $h(L)=L/2$ for a Poisson pattern.  Strong hyperuniformity ($\epsilon \ge 1$) corresponds to a large-$L$ asymptote that is constant: $h(L)=h_e$.  For this case, $\sigma_\phi^2(L)\sim \langle v\rangle/L^{d+1}$ is made dimensionally correct by the existence of $h_e$ as an emergent length rooted in the intuitive notion of what it means to be hyperuniform. Thus $h_e$ is the desired measure of structure that is independent of $L$ when the system is hyperuniform, and Eq.~(\ref{hdef}) generalizes upon this to systems with any degree of uniformity. 

%=====================
%=====================

Brief asides:  (a) The $h(L)$ spectrum is only slightly different for hyperspherical windows (see \cite{OtherWindows, DJDhudls}  and Appendix~\ref{WindowShape}). (b) It is unclear how to define a spectrum of disorder lengths from $\chi(q)$.  If the long length scale asymptotic behaviors are $h(L)\rightarrow h_e$ and $\chi(q)\rightarrow q^\epsilon$ with $\epsilon \ge 1$, then one might expect $\chi(q) \propto (qh_e)^\epsilon$ with a non-universal proportionality constant of order 1. (c) The value of $h_e$ is related to the surface coefficient $\Lambda$ \cite{TorquatoPRE2003} (see Table~\ref{hoverb} in Appendix~\ref{surfaceco}).  (d) For $d=2$ applications, as below, areas serve as the relevant $d$-dimensional volumes.

% These points are made elsewhere, without interupting the flow:
%The definitions and bounds we have presented here are all calculated from the central point (CP) representation of our configurations. We could have also chosen to treat the particles as extended objects in which case $\chi(q)$ is computed from a binary image of the system and local volume fraction depends on partial overlaps of particles with measuring windows \cite{ZacharyJSM2009, ZacharyPRL2011, WuPRE2015, ParisiPRE2017, DJDhudls}. Regardlesss of the choice of particle representation the results for the spectral density at small $q$ are unchanged. However this choice has a rather dramatic affect on the form of $\sigma_\phi^2(L)$ and consequently $h\left( L \right)$ \cite{WuPRE2015}. We use the CP representation because as we have shown it allows us to make preditctions about the form of $h\left( L \right)$ at many length scales. We currently do not know of a way to make predictions like the separated particle limit \cite{ATCpixel} and or those presented in Appendix \ref{GenTheo} from EP representation. Also  spectra from the EP representation depend on details of particle shape rather than just their spatial arrangement, making it harder to analyze features in the latter -- which is our interest.  In particular, particle-shape effects totally obscure the short- and intermediate-scale behavior we are able to see and richly interpret using the CP representation.

%------------------------------------------------------------------- SIMULATION AND ANALYSIS METHODS:

\section{Methods}

\subsection{Simulation Details}

\begin{figure*}[ht]
\includegraphics[width=7in]{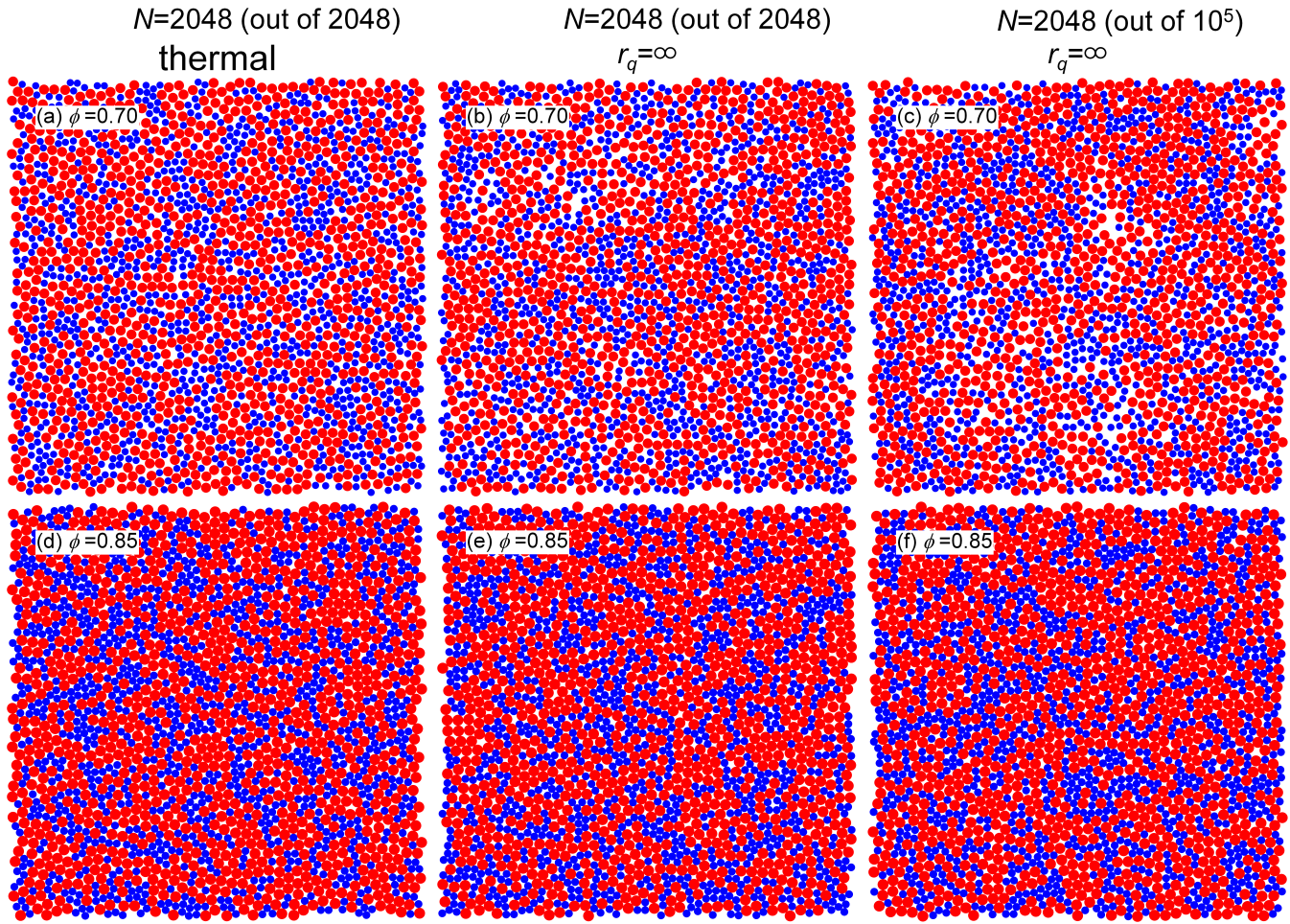}
\caption{Configurations of 2048 bidisperse disks with global packing fraction $\phi=0.70$ (a-c) or $\phi=0.85$ (d-f)  generated by different preparation protocols as labeled. Parts (a),(b),(d),(e) are the entire systems and parts (c),(f) are a subset of a system with $N=10^5$ particles. Parts (a-c) are unjammed configurations and there are noticably larger voids in (b) and (c) than in (a). Parts (e-f) are jammed configuration and they are relatively indistinguishable. The differences seen by eye are quantified with $h(L)$ because the normalized spectra are protocol and system size dependent for $\phi<\phi_c$ but collapse for the for $\phi \ge \phi_c$. }
\label{sample_configs}
\end{figure*}

Our aim is to measure and interpret $\chi(q)$ and $h(L)$ spectra for various simulated 2-dimensional particle configurations in a box with periodic boundary conditions, using the following methods.  We start with ``Einstein patterns" \cite{ATCpixel} for general intuition, and then turn our main attention to soft disks.   For {\it Einstein patterns}, monodisperse particles are close-packed on a square lattice then given a Gaussian kick with standard deviation of 1/2 lattice spacing.  These configurations match Einstein's simplified picture of crystal structure at $T>0$, and are hyperuniform by construction.  In addition, we also analyze configurations where hyperuniformity is destroyed by randomly selecting a fraction $f=10$\% of the particles and changing their areas, half by $\delta=+15\%$ and half by $-\delta$.  

For soft disk configurations we use three different protocols to equilibrate 50-50 bidisperse harmonically-repulsive disks with 7:5 diameter ratio that are initially placed at random into a box with periodic boundary conditions. Two particles $i$ and $j$ interact with a repulsive pairwise potential
\begin{equation}
	V_{ij}=\frac{\varepsilon}{2}\left(1-\frac{r_{ij}}{R_i+R_j}\right)^2 \Theta \left(1-\frac{r_{ij}}{R_i+R_j}\right)
\label{pairpot}
\end{equation}
where $r_{ij}$ is the distance between the particle centers, $R_i$ and $R_j$ are the particle radii, and $\Theta(x)$ is the Heavyside step function.  Configuration energies and temperatures are measured in units of the scale $\varepsilon$.  Times are measured in units of $\sqrt{(m\sigma^2 / \varepsilon)}$ where $m$ is particle mass, which is the same for each species, and $\sigma$ is the small particle diameter.  Energies and temperatures are measured in units of $\varepsilon$.  Forces are measured in units of $\varepsilon/\sigma$.

In the {\it thermal} protocol, {\tt LAMMPS}~\cite{PlimptonJCP1995} is used to cool the soft disks from $T_i=0.05$ (above the glass transition) to $T_f=10^{-7}$ (below the glass transition) over $5\times 10^6$ reduced time units, then to equilibrate for an additional $10^7$ time units and finally to quench to $T=0$.  The initial and final temperatures are chosen such that $T_i$ is high enough to make the initial state an equilibrium liquid and $T_f$ is far enough below the glass transition temperature for the system to be essentially free of thermally-induced rearrangements as per $T=0$ mechanical equilibrium.   A ``time step" is where an update to the molecular dynamics is performed and the thermal bath is cooled; during each time step the system is attached to an external bath using the Nose-Hoover thermostat and is kept at constant NVT.  This protocol was used in Ref.~\cite{RieserPRL16} for $N=2048$ particles at various area fractions; here, we only use those configurations.

In the  {\it quench} protocols, configurations are cooled from $T_i \rightarrow T_f$  at a controlled rate  $r_q=(T_i-T_f)/(M \Delta t)$; $M$ is a chosen number of molecular dynamics steps and $\Delta t=0.01$ is the time step in reduced units. Configurations are created with ``infinite" $(M=0)$, ``fast" $(0<M \leq10^3)$, and ``slow" $(M \geq 10^4)$ quench rates but in all cases once the systems reach their final temperature they are brought to $T=0$ using the {\tt FIRE} algorithm  \cite{BitzekPRL2006}. 

For the infinite quench protocol where $r_q=\infty$, {\tt FIRE} is implemented immediately and it brings configurations directly from $T=\infty$ to $T=0$. The $T=\infty$ configuration is one where particles are initially placed at random, and therefore $r_q=\infty$ is the only protocol where configurations are brought to $T=0$ from a state that is not an equilibrium liquid. The controlled quench protocols  equilibrate the configurations at $T_i=8\times 10^{-3}$ first and then gradually cool them to $T_f=10^{-4}$ over $M$ time steps. Similar to the thermal protocol the values of $T_i$ and $T_f$ are chosen so they are above and below the glass transition temperature, respectively. During each time step for ``fast" quenches the particles have their velocities explicitly rescaled and their positions are updated as an NVE ensemble.  During each time step for ``slow" quenches particles have the same low temperature molecular dynamics implemented in {\tt LAMMPS} as in the thermal protocol. In either case, as mentioned earlier, once the systems reach their final temperature they are brought to $T=0$ using {\tt FIRE}. We expand on our methods in the Appendix Sec.~\ref{Generation} and information about packing fraction and the number of configurations generated with particular number of particles for all protocols is collected in Table~\ref{config_info} of that section.

Example soft disk configurations for two different equilibration protocols and two different total number of particles are displayed in Fig.~\ref{sample_configs}, both below jamming $\phi=0.70$ and above jamming $\phi=0.85$.

\subsection{Analysis Details}

To compute $\sigma_\phi^2(L)$ we randomly place $L\times L$ square windows throughout the system.  The variance is deduced from the resulting set of local volume fractions, and then converted to $h(L)$ using Eq.~(\ref{hdef}).  The number $w$ of placements equals the ratio of system to window areas, constrained to $w\ge 10^2$ and either $w\le 10^4$ (for smaller systems where we average over many configurations) or $w\le 10^5$ (for larger systems where we average over 5 or fewer configurations).  The statistical uncertainty in the variance is $\Delta \sigma_\phi^2(L) = {\sigma_\phi}^2(L) \sqrt{2/(s-1)}$ where the number of independent samples is estimated by $s=[1-(1-f)^w]/f$ with $f$ being the ratio of window to sample volumes.  A correction is needed for small windows; see Section~IIIA of Ref.~\cite{ATCpixel}.

To compute $\chi(q)$ we take the Fast Fourier Transform (FFT) of a digital image where particle areas are assigned to pixel locations that correspond to particle centers.  Images are taken to have edge lengths of $\{2^{12}, 2^{13}, 2^{14}\}$ pixels for systems of $N=\{2048, 10^5, 10^6\}$ particles. These image sizes are large enough to avoid noticeable artifacts, which we demonstrate in Appendix \ref{ImiSize}.  Values of $\chi({\bf q})$ and $|{\bf q}|$ are averaged over annuli with outer radius equal to 1.1 times inner radius.  These methods may be used for experimental data \cite{experiment}.   While it is straightforward to predict the expected uncertainty in the variance and hence in $h(L)$, we are unaware of how to do so for the spectral density.

%------------------------------------------------------------------- RESULTS: 

\section{Results}

\subsection{Jammed Configurations}

\begin{figure}[ht]
\includegraphics[width=3.0in]{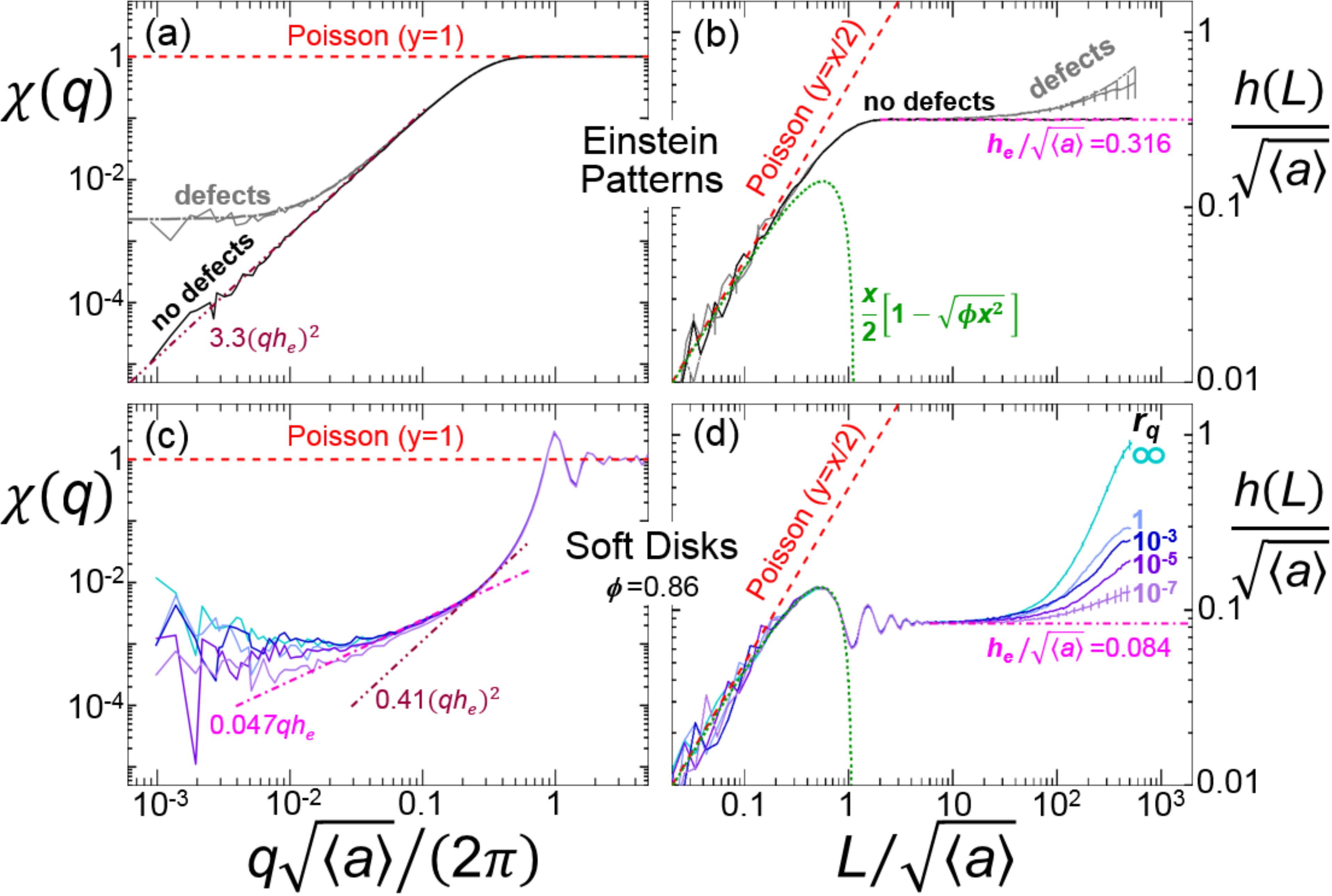}
\caption{Spectral density and hyperuniformity disorder length spectra for Einstein patterns and jammed soft disks created at different quench rates, $r_q$.  At small lengths, $h(L)$ matches the separated-particle lower bound, $L/2(1-\sqrt{\phi L^2/\langle a\rangle})$, but $\chi(q)$ deviates irregularly from the Poisson limit.  At intermediate lengths, $h(L)$ becomes constant, but for the soft disks $\chi(q)$ at first plunges precipitously then shows no such obvious signature of incipient hyperuniformity.  At long lengths, both $h(L)$ and $\chi(q)$ show that hyperuniformity is destroyed except for the defect-free Einstein patterns.  But only $h(L)$ shows a clear trend toward hyperuniformity as $r_q$ is reduced.  Putative and actual hyperuniformity are characterized by the value of $h_e$, which cannot be extracted from $\chi(q)$. }
\label{EinsteinJammed}
\end{figure}

Spectra for the Einstein patterns are displayed in Figs.~\ref{EinsteinJammed}a-b.  At large $q$ and small $L$, $\chi(q)$ and $h(L)$ approach the Poisson pattern bounds of $1$ and $L/2$, respectively.  The latter and the separated-particle limit of Eq.~(\ref{separated}) and $h(L)=(L/2)(1-\sqrt{\phi L^2/\langle a\rangle})$ hold intuitively because small windows have either zero or one particle center, at random according to $\phi$ and window size, no matter what the ordering at longer scales.  For $L>\mathcal O(\sqrt{\langle a\rangle})$ the $h(L)$ spectra quickly roll over to a constant, $h(L)\approx h_e$, with the value of $h_e$ being about one half the standard deviation of the Gaussian kicks \cite{ATCpixel}.  For the defect-free Einstein pattern, this is the true asymptotic behavior and is hyperuniform by construction.  Correspondingly, $\chi(q)\sim q^2$ is observed at small $q$.  Note that the $h(L)$ spectra are truncated beyond 1/2 system width, $L_{sys}$, where the variance is systematically suppressed; by contrast, $\chi(q)$ data extend down to $q_{min}=2\pi/L_{sys}$ without finite-size artifacts, but with blooming statistical uncertainty that is unclear how to predict.

%=====================
%=====================

For the defective Einstein patterns, spectra are seen in Figs.~\ref{EinsteinJammed}a-b to be identical to the defect-free case except for Poissonian fluctuations at long length scales.  The exact prediction in terms of the defect-free case, $h(L)=h_o(L)+\delta^2 f L/4$ (see Appendix \ref{defictiveco}), matches the data.  The corresponding signature in $\chi(q)$ is a crossover to a constant for small $q$, which can be fit to the empirical form $\chi(q)=[\chi_o(q)+c]/(1+c)$.  Altogether, the defective Einstein spectra display the key features we shall see below for packings that are nearly hyperuniform: (i) nearly random at small lengths; (ii) incipient hyperuniformity at intermediate lengths; (iii) growing fluctuations at long lengths.  Furthermore, we have rich quantitative intuition for $h(L)$, but not for $\chi(q)$, in each regime that carries over to the soft disk configurations: (i) $h(L)=(L/2)(1-\sqrt{\phi L^2/\langle a\rangle})$; (ii) $h(L)=h_e$ with $h_e$ indicating the size of particle displacements from perfect uniformity; (iii) $h(L) = \beta L$ with $\beta$ indicating with size of hyperuniformity-destroying defects.

As a first soft-disk example, spectra for $N=10^6$ particles jammed at $\phi=0.86$ by the $r_q=\infty$ protocol are shown in Figs.~\ref{EinsteinJammed}c-d.  The three generic regimes are more apparent and more clearly demarcated in $h(L)$ than in $\chi(q)$ -- random at small lengths, seemingly-hyperuniform at intermediate lengths, and defective at large lengths.  There are only a few differences from the Einstein pattern results:  First, the data veer below the $h(L)=L/2$ upper bound but remain in accord with the separated-particle lower-bound out to nearly $L=\sqrt{\langle a\rangle}$.  Therefore $\sigma_\phi^2(L) =\phi(\langle v \rangle/V_\Omega - \phi)$ can be used to determine average particle size and packing fraction for unknown samples.  Second, there are decaying oscillations in the crossover of $h(L)$ to a constant, with period set by $\sqrt{\langle a\rangle}$ as per the pair correlation function for dense disordered systems.  Oscillations are also seen in $\chi(q)$, but occur in the small-length ($q \sqrt{\langle a\rangle} \gtrsim 1$) regime; apparently, $\chi(q)=1$ is an upper bound only for small enough $q$.  Lastly, the long-range fluctuations appear to be super-Poissonian, where $h(L)$ grows faster than $L$ and $\chi(q)$ turns up at small $q$.  We focus on two quantitative features:  First, the value of $h(L)$ in the middle regime is $h_e / \sqrt{\langle a \rangle} \approx 0.084$.  This is noticeably less than the Lindemann constant ($0.15-0.30$) and only slightly larger than for close-packed disks on a square lattice (0.082 in Table~\ref{hoverb}); therefore, the intermediate range packing structure is remarkably uniform in an absolute sense.  Second, the crossover to long-range fluctuations may be characterized by a cutoff length $L_c$ beyond which $h(L)$ rises noticeably (say 10\%) above $h_e$.  For the $r_q=\infty$ spectrum in Fig.~\ref{EinsteinJammed}d, this cutoff is $L_c \approx 30\sqrt{\langle a\rangle}$.

\begin{figure}[ht]
\includegraphics[width=3 in]{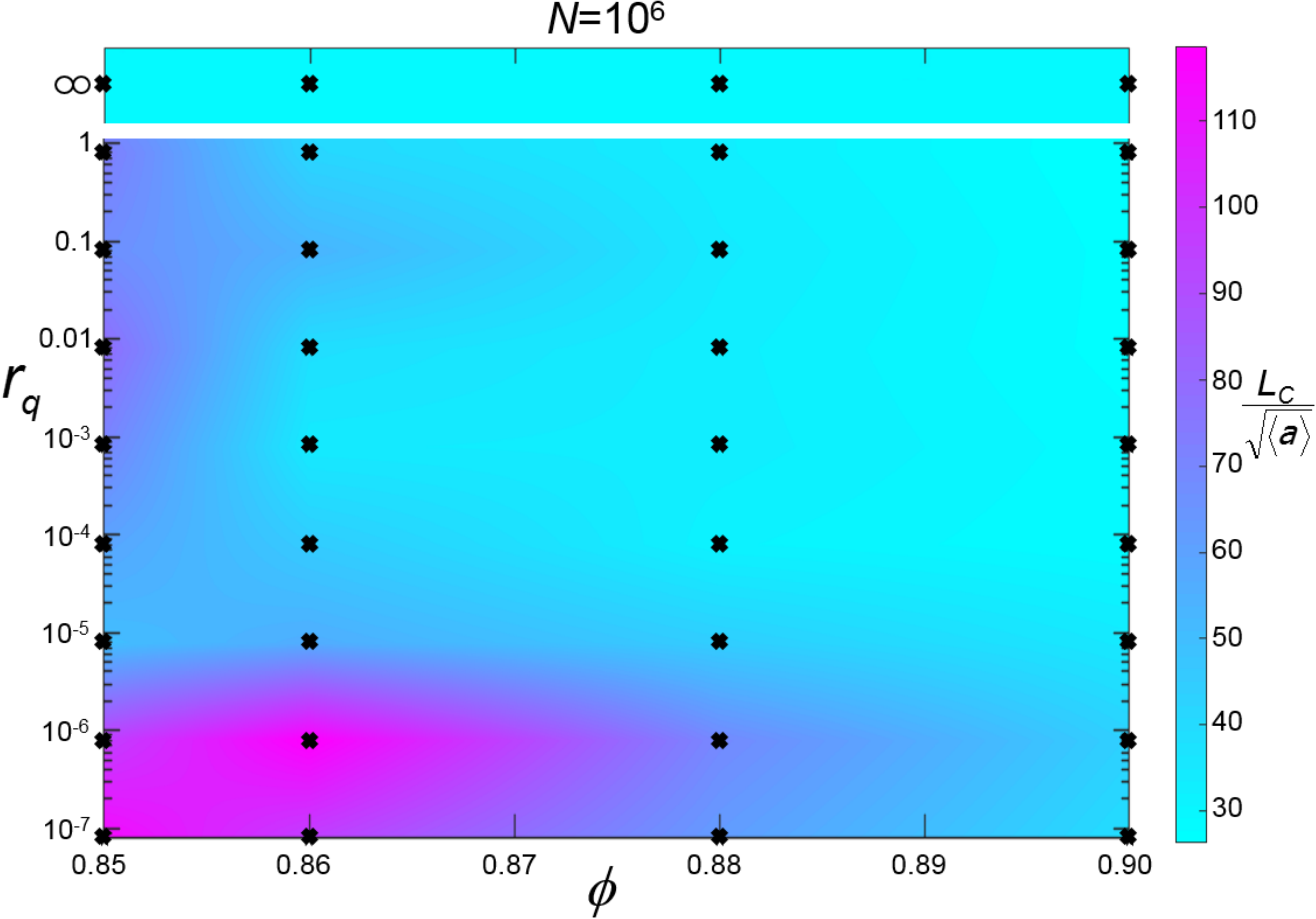}
\caption{The uniformity cutoff length $L_c$, defined by where $h(L)$ rises 10\% above $h_e$. The colorized contours interpolate between measured $L_c$ values at quench rates and volume fractions indicated by the black symbols. The data are from packings of $N=10^6$ particles.}
\label{Lc_contour}
\end{figure}

These observations raise a number of questions:  How can $h_e$ and $L_c$ be varied, what are the consequences, and what is the true long-ranged asymptotic behavior of the spectra?  To explore these issues, we begin by systematically reducing the quench rate $r_q$ at fixed $\phi$ and $N$.  The spectra data in Figs.~\ref{EinsteinJammed}c-d show that short range behavior and the value of  $h_e\approx 0.084\sqrt{\langle a\rangle}$ at intermediate scales are totally unaffected.  The only change is in the long-range structure beyond a cutoff $L_c$, which evidently increases for smaller $r_q$.  Values of $L_c$ are collected in Fig.~\ref{Lc_contour} for all packings fractions and quench rates. Strikingly for rapid quenches the value determined previously for $\phi=0.86$ of $L_c \approx 30\sqrt{\langle a\rangle}$ is the same for all rapid quenches, independent of $\phi$ and other values of $L_c$ increase only slightly for the slowest quenches we can achieve.  There is a slightly stronger increase in $L_c$ with decreasing $\phi$.  Frozen-in long-ranged density fluctuations can be partially annealed to an extent that depends on $r_q$.  However, this effect is not dramatic: even at the slowest quench rate we can achieve at present, $r_q=10^{-7}$, the value of $L_c$ is less than ten times greater than for $r_q=\infty$. It is possible that $L_c$ actually diverges, and the packings become truly hyperuniform, in the dual limit of $\phi\rightarrow\phi_c$ and $r_q\rightarrow 0$.  But if so, the divergence is slow and substantially more computational power would be required to demonstrate it convincingly. 

%Supplemental figures \cite{SoftSupp} exhibit the same behavior for other $N$ and $\phi>\phi_c$.  In particular, $L_c$ is nearly independent of these parameters unless $L_{sys}$ is smaller than $\mathcal O(3L_c)$ or $\phi$ is greater than about $\phi_c+0.04$.  

%=====================
%=====================

As an aside, the $L_c$ values in Fig.~\ref{Lc_contour} are determined based on where the spectra of $h\left( L \right)$ become larger than $(1+p) h_e$ where $p=0.1$. We choose this value of $p$ because it is large enough that $L_c$ is insensitive to statistical noise in the spectra of $h(L)$ and is also small enough that it is plainly obvious the system is not hyperuniform.  If $p$ were somewhat different, then the values of $L_c$ would change but the qualitative features in Fig.~\ref{Lc_contour} would remain the same.

\subsection{Unjammed Configurations}

%=====================
\begin{figure}[ht]
\includegraphics[width=3.0in]{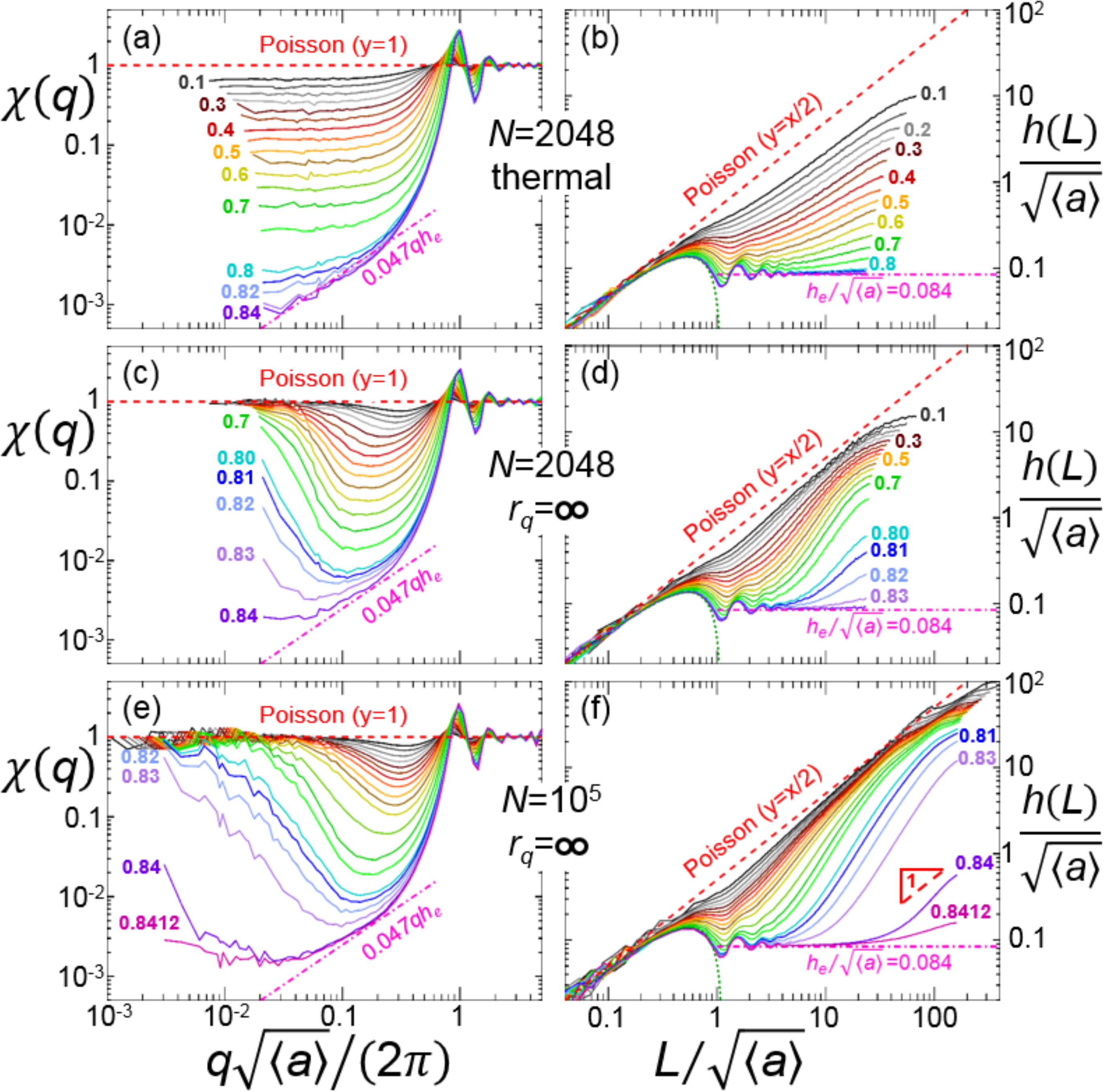}
\caption{Spectral density and hyperuniformity disorder length spectra for unjammed soft disk configuration created at various area fractions and system sizes, using two different protocols.  The sequence of area fractions is $\phi=\{0.10, 0.15, 0.20,\ldots, 0.75, 0.80, 0.81, 0.82, 0.83, 0.84\}$ from top to bottom.  In (e-f) $\phi=0.8412$ is also included.}
\label{datastacks}
\end{figure}
%=====================

We now turn to unjammed configurations, where $\chi(q)$ and $h(L)$ data are shown in Fig.~\ref{datastacks} for the thermal and $r_q=\infty$ protocols at a wide range of area fractions.  These spectra exhibit more dramatic changes versus $\phi$ than those for jammed configurations.  In the dilute limit, the data approach $\chi(q)=1$ and $h(L)=L/2$ as expected.  With increasing $\phi$, avoidance of disk-disk overlaps induces order and causes the spectra to trend downward.  For $\phi<0.75$ the thermal configurations exhibit Poissonian long-range fluctuations, as judged by the constancy of $\chi(q)$ at low $q$ and by the linearity of $h(L)$ at large $L$.  This may also happen at larger $\phi$, but even bigger systems are needed in order to investigate.  By contrast the $r_q=\infty$ configurations are distinctly super-Poissonian in that $\chi(q)$ turns up at small $q$ and $h(L)$ grows faster than $L$.  Such behavior cannot persist to arbitrarily large lengths, and indeed the spectra appear to merge with the Poisson pattern bounds of $\chi(q)=1$ and $h(L)=L/2$.  This is remarkable especially at high $\phi$: such systems appear hyperuniform at intermediate scales but {\it totally} random at long length scales!

Another feature of note in the super-Poissonian behavior of the $r_q=\infty$ spectra in Fig.~\ref{datastacks} is that it depends on system size.  In particular, the data in the bottom two rows do not coincide at intermediate lengths.  This contrasts with results at $\phi>\phi_c$ shown in Appendix  \ref{SysSize}, where larger systems exhibit spectra that overlap and extend the spectra of smaller systems.  Above jamming, bigger systems simply build out from smaller systems.  But below jamming, {\it bigger is different}.  Athermal suspensions could have structural features that are similarly extensive, and their rheology could hence be size-dependent.

The final key result in Fig.~\ref{datastacks} is perhaps the most obvious:  As $\phi$ increases toward jamming, the $h(L)$ spectra trend down at different rates and shapes to the {\it same} constant value $h_e=(0.084\pm0.001)\sqrt{\langle a\rangle}$ seen earlier.  Here the average and uncertainty are based on all spectra, for all protocols, and for all $\phi$ both above and below $\phi_c$.  Empirically, the spectral densities may trend toward $\chi(q)=0.047 q h_e$, which is close to $q h_e/(2\pi^2)$;  however, this is difficult to judge because the range over which the power law holds is much less than the range over which $h(L)\sim L^0$ holds, especially if the oscillatory range of $h(L)$ is included.   Below jamming, in Fig.~\ref{datastacks}c, the possible form of the asymptotic behavior of $\chi(q)$ and its relation to the value of $h_e$ is even less clear.

\begin{figure}[ht]
\includegraphics[width=3in]{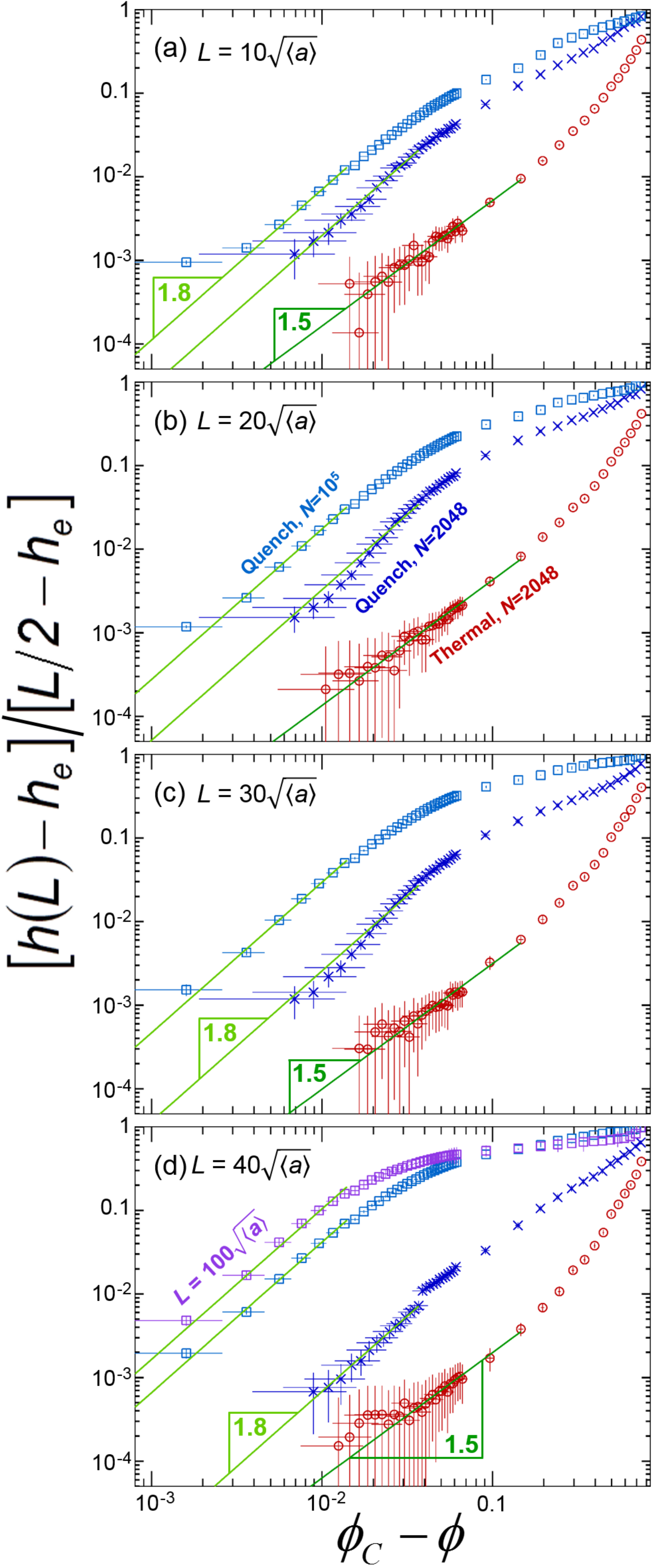}
\caption{Area fraction dependence of the hyperuniformity disorder length evaluated at various $L$ as labeled, scaled to decay from $1\rightarrow 0$ as $\phi$ goes from $0\rightarrow \phi_c$, for the same hard-disk configurations shown in Fig.~\ref{datastacks} of the main text.  The expression on the y-axis are evaluated at the $L$ labeled in black on the panel unless otherwise noted and the data have a power-law exponent of the final decay (solid lines) of 1.5 for the thermal configurations and 1.8 for the quench configurations,  independent of $L$. In parts (a-c), the curves from top to bottom have a critical packing fraction of $\phi_c=\{0.8416 \pm 0.0003, 0.8409\pm0.0012 , 0.8465\pm0.0005 \}$ and part (d) has has an addition curve in purple with $\phi_c =0.8416 \pm 0.0003$.}
\label{h_min_he_norm}
\end{figure}

The way $h(L)$ spectra trend toward $h(L)=h_e$ with increasing $\phi$ may be investigated by log-log plots showing how ${\mathcal H}(L) = [h(L)-h_e]/(L/2-h_e)$ vanishes versus $\phi_c-\phi$.  Results at various fixed $L$ are plotted in Fig.~\ref{h_min_he_norm}.  The observed difference vanishes as a power law, $(\phi_c-\phi)^{1.5}$ for the thermal protocol and $(\phi_c-\phi)^{1.8}$ for the infinite quench protocol.  These powers hold for different $L$ values and for larger $N=10^5$ systems. Both the power and proportionality constants of the power law depend on protocol but only the proportionality constants change for the larger $N=10^5$ systems (bigger is different). Once these powers and proportionality constants are known, one can use $h\left( L \right)=h_e(1+p)$ in the numerator of ${\mathcal H}(L)$ and $L=L_{sys}$ in the denominator to find how close $\phi$ needs to be to $\phi_c$ in order to create a configuration that would appear hyperuniform. It becomes increasingly difficult to generate hyperuniform configurations for larger systems because $h_e$ remains constant,  $L_{sys}$ increases and the proportionality constant grows larger. 

The origin of the power laws and the larger-is-different effect merit further study but it all supports what is seen by eye in Fig.~\ref{datastacks}:  Configurations approach hyperuniformity in the limit $\phi\rightarrow{\phi_c^-}$ for both quench protocols.  However, Ref.~\cite{Ozawa2017} studied 3D packings approaching $\phi_c^-$ and observed from $\chi(q)$ that small packings deviate more strongly from hyperuniformity with increasing $\phi_c$.  Therefore, it appears that even the limiting behavior as $\phi\rightarrow{\phi_c^-}$ may depend on protocol (and possibly dimension).  Because we find that large-scale structure is extensive below jamming, studies of system size effects using their protocols are probably necessary in order to settle this issue. 

%It is unclear whether hyperuniformity is approached in the dual limit $\phi\rightarrow{\phi_c^+}$ and $\{r_q, T_f\}\rightarrow 0$ (c.f.\ Fig.~\ref{EinsteinJammed}b); if it does, it is extremely slowly, perhaps logarithmically.  Evidently, hyperuniformity is more easily approached from below jamming. 
%If it is achieved, then the asymptotic scaling is specified by $h_e=0.084\sqrt{\langle a\rangle}$ independent of protocol.

%=====================
%=====================

\section{Conclusion}

In conclusion, we have defined an emergent length $h(L)$ and have shown how to compute its statistical uncertainty as well as upper/lower bounds and expectations at short, intermediate, and long length scales.  We call this analysis method ``hyperuniformity disorder length spectroscopy" (HUDLS).  Unlike scaling tests for hyperuniformity, which focus only on large-$L$ asymptotics, HUDLS brings rich physically meaning to the {\it value} of the spectrum at {\it all} $L$.  Furthermore, as summarized in the caption of Fig.~\ref{EinsteinJammed}, features in $h(L)$ at short, intermediate, and long $L$ are far easier to identify and quantitatively interpret than in $\chi(q)$ vs $q$.  As applied to soft disks, configurations become more uniform for slower quench rates and as $\phi\rightarrow \phi_c$ from both above and below, with unexpected system-size- and protocol-dependent long-range structural features.  This sharpens the challenge of how to identify and mitigate the subtle structural defects that destroy hyperuniformity at long length scales.  Rattlers cannot be responsible, since their number decreases with distance above jamming \cite{Epitome} and with increasing $r_q$~\cite{Ozawa2017}, opposite to $L_c$ trends.  We speculate that under-packed regions defined by the power-law tail in the $Q_k$ metric \cite{RieserPRL16} play a role.  It would be interesting to establish some such connection and to explore the consequences for rheological behavior.  Furthermore, it would be interesting to explore how large $L_c$ must be, and perhaps how small $h_e$ must be, in order to endow a material with the special properties associated with true hyperuniformity. The hyperuniformity disorder length concept can also be extended to describe fluctuations in quantities other than density, such as local topology or connectivity~\cite{Roth2013, Hexner2017}.  Altogether, HUDLS thus offers a general and intuitive real-space method to characterize the spectrum of structural features as a fundamental step in understanding materials properties.

%------------------------------------------------------------------- END PAPER:

\begin{acknowledgments}
We thank C. Goodrich for data from Ref.~\cite{RieserPRL16}, and L. Berthier, R. Dreyfus, S. Teitel and S. Torquato for helpful conversations.  We acknowledge NASA grant NNX14AM99G (DJD); NSF grants MRSEC/DMR-1720530 (AJL, DJD) and DMR-1305199 (DJD); National Natural Science Foundation of China grants 21325418 and 11574278 (NX); Fundamental Research Funds for the Central Universities grant 2030020028 (NX).
\end{acknowledgments}

%------------------------------------------------------------------- BEGIN APPENDIXES

\appendix

\section{General Theory} \label{GenTheo}

\subsection{Periodic Lattices of Close-Packed Spherical Particles}\label{surfaceco}

Perfect crystals are hyperuniform and have a constant hyperuniformity disorder length, $h_e$, whose value is characteristic of the particular lattice.  For infinite periodic lattices of spacing $b$, Ref.~\cite{TorquatoPRE2003} defined a surface coefficient $\Lambda$ in terms of the number variance for spherical windows of radius $R$ by $\sigma_N^2(R)=\Lambda(R/b)^{d-1}$.  Since lattices are spatially anisotropic, spherical windows are needed to avoid oscillations in the variance (e.g. see Section II.C of Ref.\cite{ATCpixel}).  Results were given for special lattices in 2-3 dimensions in Tables~III-IV, respectively, in terms of the volume fraction $\phi$ of the corresponding close packing of particles with radius $r=b/2$.  This can be related to $h_e$ by comparing with its definition for large windows (any shape):  $\sigma_N^2(R)=\rho A_\Omega h_e$ where $\rho=\phi/v$ is number density, $v$ is particle volume, and $A_\Omega$ is the window surface area.  In two dimensions, where $v=a=\pi r^2$ and $A_\Omega=2\pi R$, this gives $h_e/b = \Lambda /(8\phi)$.  In three dimensions, where $v=(4/3)\pi r^3$ and $A_\Omega=4\pi R^2$, the correspondence is $h_e/b = \Lambda /(24 \phi)$.  Resulting values of $h_e/b$ are given in Table~\ref{hoverb}.  The ordering of the lattices in the table is the same as in Ref.~\cite{TorquatoPRE2003}, which is monotonic in scaled $\Lambda$ and also in $h_e\rho^{1/d}$.

%=====================
\begingroup
\captionsetup{justification=raggedright,singlelinecheck=false}
\setlength{\tabcolsep}{8pt}
\renewcommand{\arraystretch}{1.3}
\begin{table*}[ht]
\begin{center}
\begin{tabular}{l r c c c c c c } 
\hline
\hline
\textbf{Lattice} & \boldmath $\phi$\ \ \ \ \ \ \ \ \ \ \  & \boldmath $\Lambda / \phi^{1/2}$ & \boldmath$\Lambda / \phi^{2/3}$  &  \boldmath$h_e / b$ & \boldmath$h_e \rho^{1/d}$ & \boldmath $h_e / a^{1/2}$ & \boldmath $h_e / v^{1/3}$   \\
\hline
triangular & $\pi / \sqrt{12} \approx 0.906900$ & 0.508347 &     & 0.0667253 & 0.0717010 & 0.0752915 &  \\ 
square     & $\pi / 4              \approx 0.785398$ & 0.516401 &     & 0.0728370 & 0.0728370 & 0.0821878 &  \\ 
honeycomb & $\pi / \sqrt{27}  \approx 0.604600$ & 0.567026 &     & 0.0911547 & 0.0799775 & 0.1028570 &  \\ 
Kagome & $\pi \sqrt{3/64}  \approx 0.680175$ & 0.586990 &     & 0.0889673 & 0.0827934 & 0.1003889 &  \\ [12pt]

bcc      & $\pi \sqrt{3/64}  \approx 0.680175$ &   & 1.24476     & 0.0589749 & 0.0643490 &   & 0.0731703  \\ 
fcc       & $\pi / \sqrt{18}     \approx 0.740480$ &   &  1.24552    & 0.0573634 & 0.0643882 &   & 0.0711708  \\ 
hcp      & $\pi / \sqrt{18}     \approx 0.740480$ &   & 1.24569     & 0.0573712 & 0.0643970 &   & 0.0711805 \\ 
sc        & $\pi / 6                  \approx 0.523599$ &   &  1.28920    & 0.0666463 & 0.0666463 &   & 0.0826882 \\ 
\hline
\hline
\end{tabular}
\end{center}
\caption[l]{Close-packing volume fractions, scaled surface coefficients as given in Ref.~\protect{\cite{TorquatoPRE2003}}, and the corresponding hyperuniformity disorder lengths $h_e$ for spherical particles of radius $r=b/2$ on special lattices in $d=2$ and $d=3$ dimensions.  Values are scaled according to lattice spacing $b=2r$, number density $\rho$, and either particle area $a=\pi r^2$ or particle volume $v=(4/3)\pi r^3$. }
\label{hoverb}
\end{table*}
\endgroup

%=====================

For comparison with findings in the main text, Table~\ref{hoverb} also lists $h_e$ values scaled by the square-root of particle area in $d=2$ and by the cube-root of particle volume in $d=3$.  In particular, the simulation result $h_e/\sqrt{\langle a\rangle}=0.084$ for jammed bidisperse disks is only slightly greater than the tabulated result $h_e/\sqrt{a}=0.082$ for monodisperse disks close-packed on a square lattice.

%------------------------------------------

\subsection{Defective Configurations}\label{defictiveco}

Here we consider a central-point representation where a given configuration of particles, of volume $v$, is altered by adding a volume $\delta \times v$ to a randomly-chosen fraction $f$ of the particles.  For example, $\delta=-1$ corresponds to vacancy defects and $\delta=+1$ corresponds to double-occupancy defects.  The volume fraction variance of the defective configuration may be computed exactly in terms of that for the original configuration, $\sigma_{\phi_o}^2(L)$, as follows.  First, the local volume fraction is $\phi_\Omega=[N_o v + N_d \delta v]/V_\Omega$ where $N_o$ and $N_d$ are the number of original particle centers and introduced defects, respectively, enclosed by a given local window of volume $V_\Omega$.  So the true volume fraction changes to
\begin{eqnarray}
	\phi \equiv \overline {\phi_\Omega} &=& (\overline{N}\!_o + \delta \overline{N}\!_d)\frac{v}{V_\Omega}, \\
	&=& (1+\delta f)\phi_o, \label{phidf}
\end{eqnarray}
since the average number of particles is $\overline{N}\!_o = \phi_o V_\Omega/v$ and the average number of defects is $\overline{N}\!_d = f \overline{N}\!_o$.  Next, the volume fraction variance is

\begin{align}
	\sigma_\phi^2(L) &= \overline{ \phi_\Omega^2} - {\overline {\phi_\Omega}}^2, \\
	\begin{split}
		&=  \Big[ \left(\overline{N_o^2}-{\overline{N}\!_o}^2\right) +2\delta\left(\overline{N_oN_d}-\overline N\!_o\overline N\!_d\right) \\
		&{} \qquad + \delta^2\left(\overline{N_d^2}-{\overline{N}\!_d}^2\right) \Big]  \left(\frac{v}{V_\Omega}\right)^2, 
	\end{split}
	\\
	&= \left[ \sigma_{N_o}^2 + 2\delta \sigma_{N_oN_d} + \delta^2 \sigma_{N_d}^2 \right]  \left(\frac{v}{V_\Omega}\right)^2. \label{varN}
\end{align}

To find the center-defect covariance and the defect variance, we write $\overline{N}_o = \sum{n p_n}$ and $\overline{N_o^2} = \sum{n^2 p_n}$ where $p_n$ is the probability of finding $n$ particles in a randomly placed window.  If there are $n$ particles in a window, then the probability for $k$ of them to be defective is given by the binomial distribution as $q_k = \{n!/[k!(n-k)!]\}f^k (1-f)^{n-k}$.  Direct summation gives 
\begin{eqnarray}
	\overline{N}_d &=& \sum_n\sum_{k=0}^n(kq_k)p_n = f \overline{N}_o, \\
	\overline{N_d^2} &=& \sum_n\sum_{k=0}^n(k^2q_k)p_n = f^2\overline{N_o^2} + f(1-f)\overline{N}_o, \\
	\overline{N_oN_d} &=& \sum_n\sum_{k=0}^n(np_n)(kq_k) = f\overline{N_o^2},
\end{eqnarray}
without any need to know $p_n$.   The covariance and variance are thus
\begin{eqnarray}
	\sigma_{N_oN_d} &=& f \sigma_{N_o}^2, \\
	\sigma_{N_d}^2 &=& f^2 \sigma_{N_o}^2 + f(1-f)\overline{N_o}.
\end{eqnarray}
Plugging these expressions into Eq.~(\ref{varN}) and dividing by Eq.~(\ref{phidf}) gives the final result for the volume fraction variance of the defective configuration as
\begin{equation}
	\frac{\sigma_\phi^2(L)}{\phi} = (1+\delta f)\frac{\sigma_{\phi_o}^2(L)}{\phi_o}+\frac{\delta^2f(1-f)}{1+\delta f}\frac{v}{V_\Omega}.
\label{deltavar}
\end{equation}
If the initial configuration is hyperuniform, then the $1/V_\Omega$ Poissonian term will dominate at large windows; therefore, the defective configuration is not  hyperuniformity.  For the case of vacancies, $\delta=-1$,  Eq.~(\ref{deltavar}) reduces to a prior result computed and tested by simulation in Ref.~\cite{ATCpixel}.

We now generalize further by considering a distribution of uncorrelated defects $\delta_i$ that occur at random with probabilities $f_i$.  Here an important new ingredient is that the defect-defect covariance is computed to be $\sigma_{N_iN_j}=f_if_j\sigma_{N_o}^2 - f_if_j\overline{N_o}$, where the first term reflects the variability in number of particles per window and the second reflects multinomial statistics.  This leads to
\begin{equation}
	\frac{\sigma_\phi^2(L)}{\phi} = (1+\langle \delta\rangle f)\frac{\sigma_{\phi_o}^2(L)}{\phi_o} 
	    + \frac{\langle\delta^2\rangle f -\langle\delta\rangle^2 f^2}{1+\langle \delta\rangle f}\frac{v}{V_\Omega},
\label{deltavargen}
\end{equation}
where $\phi = (1+\langle \delta\rangle f)\phi_o$, $\langle \delta^n\rangle = (\sum \delta_i^n f_i)/\sum f_i$ and $f=\sum f_i$.

Specializing to hypercubic windows, $V_\Omega = L^d$, we now define a hyperuniformity disorder length for the defective configuration by $\sigma_\phi^2(L)/\phi = (v/L^d)\left\{ 1 - [1-2h(L)/L]^d\right\}$, and similarly for $h_o(L)$ in the initial configuration.  Solving for $h(L)$ gives
\begin{align}
	\begin{split}
	h(L) &= \frac{L}{2}-\frac{L}{2}\Bigg\{ (1+\langle \delta\rangle f)\left[1-\frac{2h_o(L)}{L}\right]^d   \\
	                   &{} \hspace{1.1in} \quad  - \frac{ f[\langle \delta\rangle +\langle \delta^2\rangle]}{1+\langle \delta\rangle f} \Bigg\}^{1/d}, \label{hdefects} 
	\end{split}
	\\
	        &\approx \left( 1 + \langle\delta\rangle f \right)h_o(L)  +    \frac{\langle\delta^2\rangle f -\langle\delta\rangle^2 f^2}{1+\langle \delta\rangle f}\frac{L}{2d}, \label{hdefectsapprox}
\end{align}
where the approximate expression is for $h\ll L$, where $\sigma_\phi^2(L)/\phi \approx 2d(vh)/L^{d+1}$ holds.  For small $L$, the spectrum is slightly shifted due to defects.  For large $L$ the effect is more dramatic: the spectrum behaves as $h(L)\propto L$, indicative of Poissonian fluctuations.  Note that these fluctuations are not due to particles that are literally within a distance $h$ of the window surface.  Rather, they are due to particles and defects in a subregion of measure $\mathcal O (L^{d-1}h)$ that is spread uniformly throughout the window.

As a specific example, if unbound defect pairs $\delta_1=-\delta_2$ occur with equal probability $f_1=f_2=f/2$, then $\langle \delta\rangle = 0$, $\langle \delta^2\rangle=\delta_1^2$, the volume fraction is unchanged, the variance is inflated over that of the original configuration by $\langle \delta^2\rangle f v / V_\Omega$, and Eq.~(\ref{hdefectsapprox}) simplifies to $h(L)=h_o(L)+\langle \delta^2\rangle f L/(2d)$.  See Fig.~2b of the main text for comparison with simulation results for such a pattern.

%------------------------------------------

\subsection{Interpretation of Small Poissonian Asymptotics for Cubic Windows}

For a totally-random Poisson pattern, the relative variance is $\sigma_\phi^2(L)/\phi = \langle v\rangle / L^d$ and the corresponding hyperuniformity length is $h(L)=L/2$.  The actual real-space spectra for a given disordered configuration may also be found to exhibit long-range fluctuations that are Poissonian, though smaller than for a totally-random pattern, for a variety of reasons.  In light of Eq.~(2) from the main text, this may be written in two equivalent ways in terms of a dimensionless number $\beta \ll 1$ that serves as an index for the strength of the small Poissonian fluctuations:
\begin{eqnarray}
	h(L) &\rightarrow& \beta L, \label{betadef} \\
	{\sigma_\phi^2(L)}/{\phi} &\rightarrow& 2d\beta\langle v\rangle / L^d.
\end{eqnarray}
For example, if $h(L)$ is found to be $h(L)=h_e[1+(L/L_C)(1+kL)]+\mathcal O(L^3)$ for $L$ much larger than the particle size, then the index is $\beta=h_e/L_C$.  One could then roughly say that the system appears hyperuniform for $L\ll L_C$ and Poissonian for $L \gg L_C$.  Though common practice,  the former is technically a misnomer since the property of hyperuniformity refers to the large-$L$ asymptotic behavior.  [N.B. Here $L_C$ is ten times greater than the cutoff $L_c$ used in the main text.]

We have presented data as well as two particular calculations in which hyperuniformity is ruined in this manner at long length scales.  Comparing the latter with the Eq.~(\ref{betadef}) definition of $\beta$ gives the connection as

\begin{numcases}
   { \beta  = \frac{1}{2d} }
	\frac{ \langle \delta v^2/v\rangle}{\langle v\rangle} \qquad \quad \substack{\text{particle-volume} \\ \text{uncertainty}}, \label{betadv} \\
	 \frac{\langle\delta^2\rangle f -\langle\delta\rangle^2 f^2}{1+\langle \delta\rangle f}  \quad {\rm defects.} \label{betadefects}
\end{numcases}
For data where the source of $h(L)=\beta L$ Poissonian fluctuations is unknown, the observed value of $\beta$ may be compared intuitively in terms of corresponding values for $\langle \delta v^2/v\rangle$ or $\{ \langle \delta^n\rangle = \sum \delta_i^n f_i/\sum f_i, f = \sum f_i\}$.

%-------------------------------------------------------------------------------------------------------------------------------------------------------------------------------------------

%------------------------------------------
\section{Simulation Details} \label{Generation}

\begingroup
\captionsetup{justification=raggedright,singlelinecheck=false}
\setlength{\tabcolsep}{12pt}
\renewcommand{\arraystretch}{3}
\begin{table*}[ht]
\begin{center}
\begin{tabular}{c c c c c } 
\hline
\hline 
\textbf{Protocol} & \boldmath${N_p}$ & \boldmath$\phi$ & \boldmath$\left< \phi_C \right> \pm \Delta \phi_C$  &  \boldmath$N_C$  \\
\hline
thermal & 2048 & \makecell{0.10, 0.15, 0.20, ... 0.70, 0.75, 0.80, \\ 0.81, 0.82, 0.83, 0.84} & $0.8465\pm0.0005$  & 200 \\ 
\hline
\multirow{3}{*}{$r_q=\infty$} & 2048 & \makecell{0.10, 0.15, 0.20, ... 0.70, 0.75, 0.80, \\ 0.81, 0.82, 0.83, 0.84}  & $0.8409\pm0.0012$  & 200 \\
& $10^5$ & \makecell{0.10, 0.15, 0.20, ... 0.70, 0.75\\ 0.78, 0.782, 0.784 ... 0.84, 0.8412, \\ 0.842, 0.844 ... 0.898, 0.90}  & $ 0.8416 \pm 0.0003$  & 5 \\
 & $10^6$ & 0.85, 0.86, 0.88, 0.90  & $<0.85$ & 1 \\
\hline
 $r_q=0.79 \times$ \makecell[l]{$\{1, 0.1, 0.01, 10^{-3}, $ \\ ~$10^{-4}, 10^{-5}, 10^{-6}, 10^{-7}\}$} & $10^6$ &  0.85, 0.86, 0.88, 0.90 & $<0.85$  & 1 \\ [3pt] 
\hline
\hline
\end{tabular}
\end{center}
\caption[l]{The number of particles, packing fraction, critical packing fraction for jamming with its uncertainty (where known) and the number of configurations generated for each packing fraction for a given protocol.  All data shown in the main paper and appendix for any particular protocol are averaged over all configurations for a given value of $\phi$.  }
\label{config_info}
\end{table*}
\endgroup

\subsection{Choice of Initial and Final Temperatures} \label{SimTemps}

For the thermal configurations the  initial and final temperatures are chosen such that $T_i$ is high enough to make the initial state an equilibrium liquid and $T_f$ is far enough below the glass transition temperature for the system to be essentially free of thermally-induced rearrangements as per $T=0$ mechanical equilibrium. We only analyze packings outside of the glass transition so the actual value of $T_g$ should not affect our analysis.

We only study systems above the jamming transition for the controlled quench protocol. For the controlled quenches $(r_q<\infty)$ the initial temperature $T_i$ and a final temperature $T_f$ are also chosen such that they are above and below the glass transition tempertaure $T_g$, respectively. Starting above $T_g$ allows particles to rearrange so the final configurations are independent of the initial particle placement. Ending below $T_g$ essentially arrests particle motion and the particles become caged. This is important because it minimizes the amount of movement implemented by the {\tt FIRE} algorithm for configurations with many molecular dynamic steps. The glass transition temperature is estimated by fitting the temperature dependence of the relaxation time of supercooled liquids to the Vogel-Fulcher function and the the value of $T_g$ corresponds to the divergence of the relaxation time. We find for $\phi=\{0.85,0.86,0.88,0.90\}$ that $T_g=\{0.000198,0.000296,0.000525,0.000726\}$ thus confirming that $T_i$ and $T_f$ are indeed above and below $T_g$ regardless of the packing fraction of the configuration.

We study systems at temperatures below $T_g$ because it is known that properties of glassy states such as potential energy, glass stability, etc depend on quench rate; we investigate whether global positional order also has such a dependence. This question is novel so we investigate $r_q$ over many orders of magnitude to see which, if any, will change the uniformity of the configurations. From the spectra of $h(L)$ and the values of $L_C$ reported in the main paper as well as Fig.~\ref{SpectralStack} shown in this appendix, we find that configurations brought to their final state using the {\tt FIRE} algorithm have the same onset of Poissonian fluctuations if they are generated with $0 \leq M \leq 10^4$ molecular dynamic steps. This new result instructs future simulations because we show that one must run at least $10^5$ MD steps before quenching otherwise the same order is acheived as quenching from an equilibrium liquid or a completely random configuration.

%------------------------------------------

\subsection{Finding the Jamming Transition} \label{ConfigDetails}

Packings are generated either far below or far above the jamming transition but $\phi_c$ is not a universal constant. While $\phi_c$ depends on the preparation protocol and can fluctuate for finite systems, it is almost always in a narrow range and coincides with random close packing; we expect this is true of our systems as well. The $N=2048$  configurations are borrowed from previous work where  $\phi_c$ is independently calculated \cite{RieserPRL16}. Using methods described in \cite{OHernPRL2002,OHernPRE2003} the researchers find that $\phi_c$ for the thermal and inifinte quench protocols are $\phi_c=\{ 0.8465\pm0.0005, 0.8409\pm0.0012\}$, respectively. 

\begin{figure}[ht]
\captionsetup{justification=raggedright,singlelinecheck=false,margin=20pt,font=small}
\includegraphics[width=3 in]{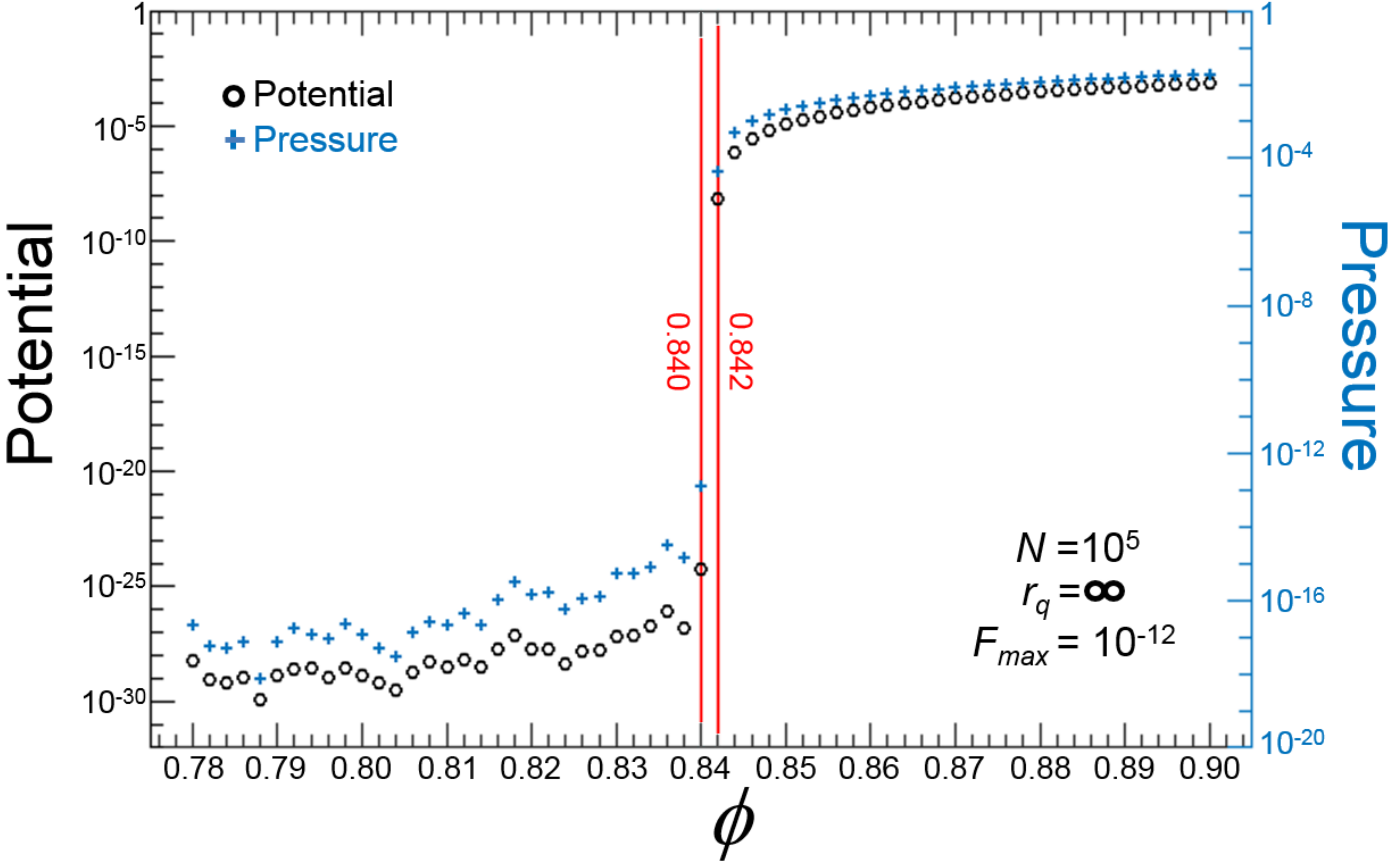}
\caption{Potential energy (left axis) and pressure (right axis) versus packing fraction for many configurations below, near and above the jamming transition. The configurations are generated with the infinite quench protocol and have $N=10^5$ particles. The values of potential energy (black circles) and pressure (blue crosses) also depend on the maximum allowed force for any given particle which is given as $F_{max}$. There is a discontinuity in both the potential and pressure between $0.840 \leq \phi \leq 0.842$ (red lines) and power law fits give an estimate for the critical packing fraction as $\phi_c=0.8416 \pm 0.0003$.   }
\label{pot_pres_phi}
\end{figure}

We generate 5 configurations for the $r_q=\infty$ protocol with $N=10^5$ particles. There are too few packings at each $\phi$ to use the same methods in determining $\phi_c$ as we did for the $N=2048$ configurations. Instead we look to the values of the potential energy and pressure for each configuration and we plot the average value of these quantities versus $\phi$ in Fig.~\ref{pot_pres_phi}. There is a clear discontinuity between $0.840 \leq \phi \leq 0.842$ for both measures and these values provide an upper and lower bound on $\phi_c$.  Below jamming, the potential energy should vanish of course; however, in practice, the simulations are stopped when the residual force on each particle falls below a very small value (see Section~IIIA, below).  To estimate $\phi_c$, a power law $\left( \phi - \phi_c \right)^\nu$ is fit to the configurational potential energy as it approaches ${\phi_c}^+$ and these fits return $\phi_c=0.8416 \pm 0.0003$. This number is similar to the value of $\phi_c$ from an earlier study by Vagberg {\it et~al.} where they use a similar protocol to ours and find that $\phi_c \simeq 0.842$ \cite{VagbergPRE2011}. All of the other quench systems we study have $\phi_c<0.85$ but only packings with $\phi\geq 0.85$ are examined which ensures they are jammed. Values and estimates of $\phi_c$ and other relevant details about all  of our configurations are presented in Table \ref{config_info}.

%=====================
%=====================

%--------------------------------------------------------------------------------------------------------------------------------------------------------------------------------------------

\section{Analysis of all of our $N=10^6$ Soft Disk Packings with $\phi>\phi_C$}

\begin{figure}[ht]
\captionsetup{justification=raggedright,singlelinecheck=false,margin=20pt,font=small}
\includegraphics[width=3 in]{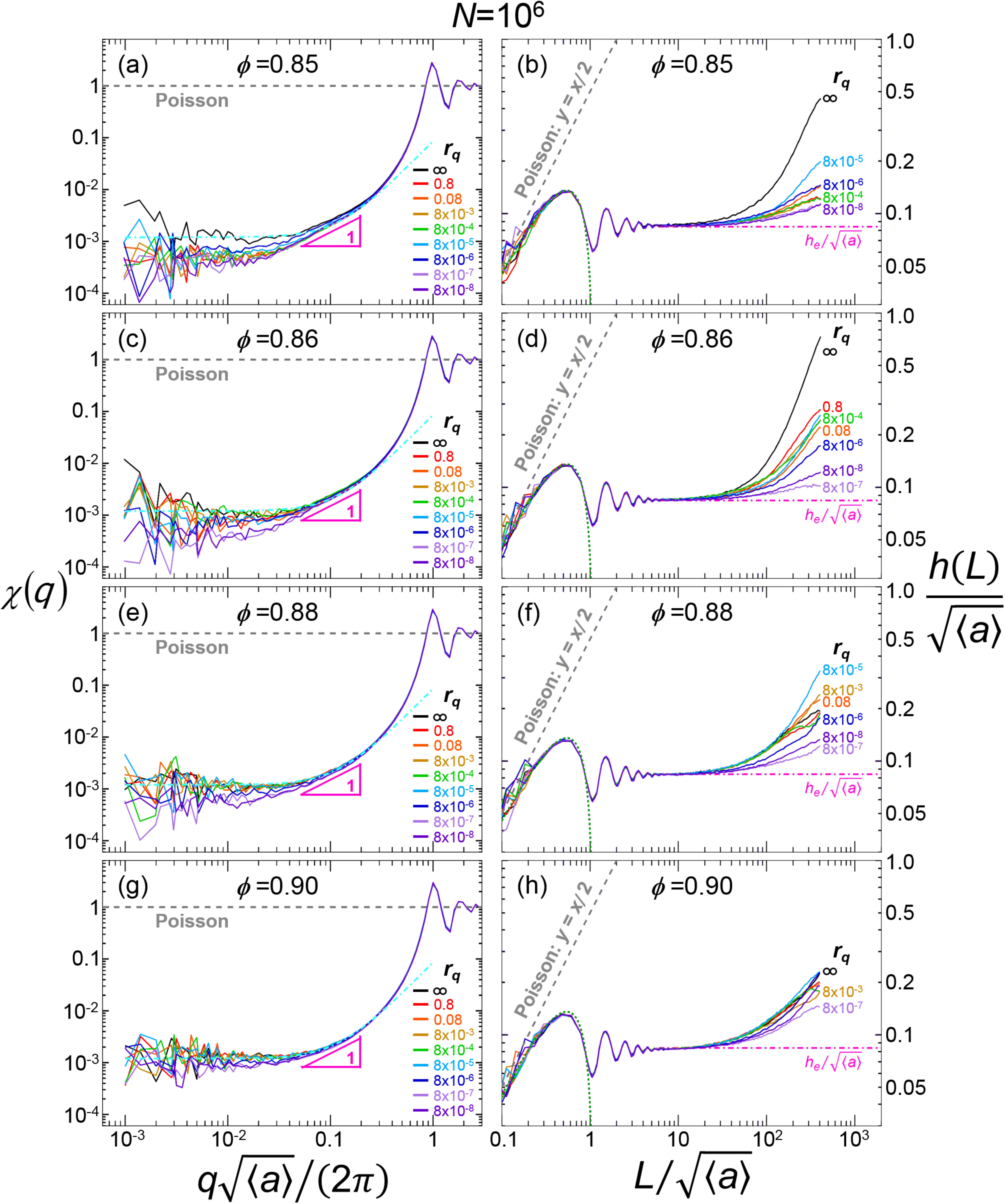}
\caption{ The spectral density versus wavevector (left panels) and the hyperuniformity length versus measuring window side length (right panels), for global packing fraction as labeled. The data are obtained from $N=10^6$ particle packings for all of the quench rates we study. The spectral density plots have a magenta triangle which demonstrates linear behavior in $q$ and a light blue dot-dashed curve $\chi \left( q \right) = \chi_o + \kappa q^2$ where $\chi_o=0.0012 \pm 0.0002$ and $\kappa = 0.084 \pm 0.003$, which was obtained by fit to the average of the $\phi=0.90$ data. In plots of the hyperuniformity disorder length some of the curves are labeled by their quench rate and the color code is the same as in the plots for the spectral density. The green dotted curve in these plots is the separated-particle lower bound, $h(L)=(L/2)(1-\sqrt{\phi L^2/\langle a\rangle})$. In all plots the gray dashed lines labeled ``Poisson" represent the expectation for a completely random arrangement of particles. }
\label{SpectralStack}
\end{figure}

Previously we showed analysis for a select packing fraction and a few quench rates but we have generated many more than what are shown. We presented data from all the packings we generate in Fig.~\ref{Lc_contour}. In Fig.~\ref{SpectralStack} we show the same comparison between $\chi \left( q \right)$ and $h\left(L\right)$ as in Fig.~2 (c),(d) in the main text but for all the configurations we generate with $N=10^6$ particles. Parts (c),(d) repeat some of the data from earlier and we can see regardless of quench rate or packing fraction that the behavior is generally the same as what was described there. 

A keen observer will notice that the data are not monotonic with quench rate for both $\chi(q)$ and $h(L)$. We suspect that this is a consequence of having only one configruation for each quench rate. Recall that at the end of each simulation a configurations is set to a local minimum with the {\tt FIRE} algorithm; it is almost certain that whatever potential minimum the configuration reaches is not the global minimum and may not even be a particularly deep local minimum. It is likely that if we had an ensemble average of many configurations at each quench rate that the curves would be monotonic. However, we can still see that it is generally true that a system closer to $\phi_c$ and with a slower quench rate is more uniform. 

This increase in order is evident in both the spectral density and the hyperuniformity disorder length spectra. For the spectral density we fit the function $\chi(q) = \chi_o + \kappa q^2$ to the $\phi=0.90$ data and use this as a fiduciary curve in parts (a,c,e,g) of Fig.~\ref{SpectralStack}.  As $\phi$ decreases towards jamming there is an average decrease in the plateau value of the spectral curves indicating an increase in order.  We note that a linear form, $\chi(q)=\chi_o+\kappa q$, does not fit the data as well.

There is a similar increase in order observed in the hyperuniformity disorder length spectra but none of the soft disk configurations we generate above jamming are hyperuniform. If they were the $h(L)$ spectra in Fig.~\ref{SpectralStack} would remain at the constant value $h(L)=h_e$ for all large $L$. The spectra deviate from this constant and the systems are ultimately deemed not hyperuniform. The systems still have some degree of order and are uniform out to a particular length scale. To quantify the general level of uniformity, we measure a cut-off length $L_c$ above which $h(L)$ rises 10\% above $h_e$ (i.e.\ below which the system could be mistaken as hyperuniform).  The results collected in Fig.~\ref{Lc_contour} show that $L_c$ is about $30\sqrt{\langle a\rangle}$ for rapid quenches, independent of $\phi$, and increases by only slightly for the slowest quenches we can achieve.  There is a slightly stronger increase in $L_c$ with decreasing $\phi$.  It's possible that $L_c$ actually diverges, and the packings become truly hyperuniform, in the dual limit of $\phi\rightarrow\phi_c$ and $r_q\rightarrow 0$.  But if so, the divergence is slow and substantially more computational power would be required to demonstrate it convincingly.  Apparently, hyperuniformity is more easily approached from below jamming.

%=====================
%=====================

%------------------------------------------
\section{Testing our Methods}

\subsection{Influence of Equilibration Precision}

It is shown in the main text that large jammed packings of bidisperse soft disks, $\phi>\phi_c$, have an upturn of $h(L)$ at large $L$ to Poissonian behavior, and therefore are not hyperuniform.  Here we check whether this could be an artifact that arises from poor equilibration -- in other words, from imprecision in which the net force on each particle is brought to zero by a given numerical quench method.  To investigate, we run the simulations until the net force on each particle in the systems falls below a maximum allowed force residue, $F_{max}$.  Data in the main text are for $F_{max}=10^{-12}$.  Here we generate additional $r_q=\infty$ packings with $N=10^5$ particles for $F_{max}=\{10^{-4},10^{-8},10^{-16}\}$.  

\begin{figure}[ht]
\captionsetup{justification=raggedright,singlelinecheck=false,margin=20pt,font=small}
\includegraphics[width=3 in]{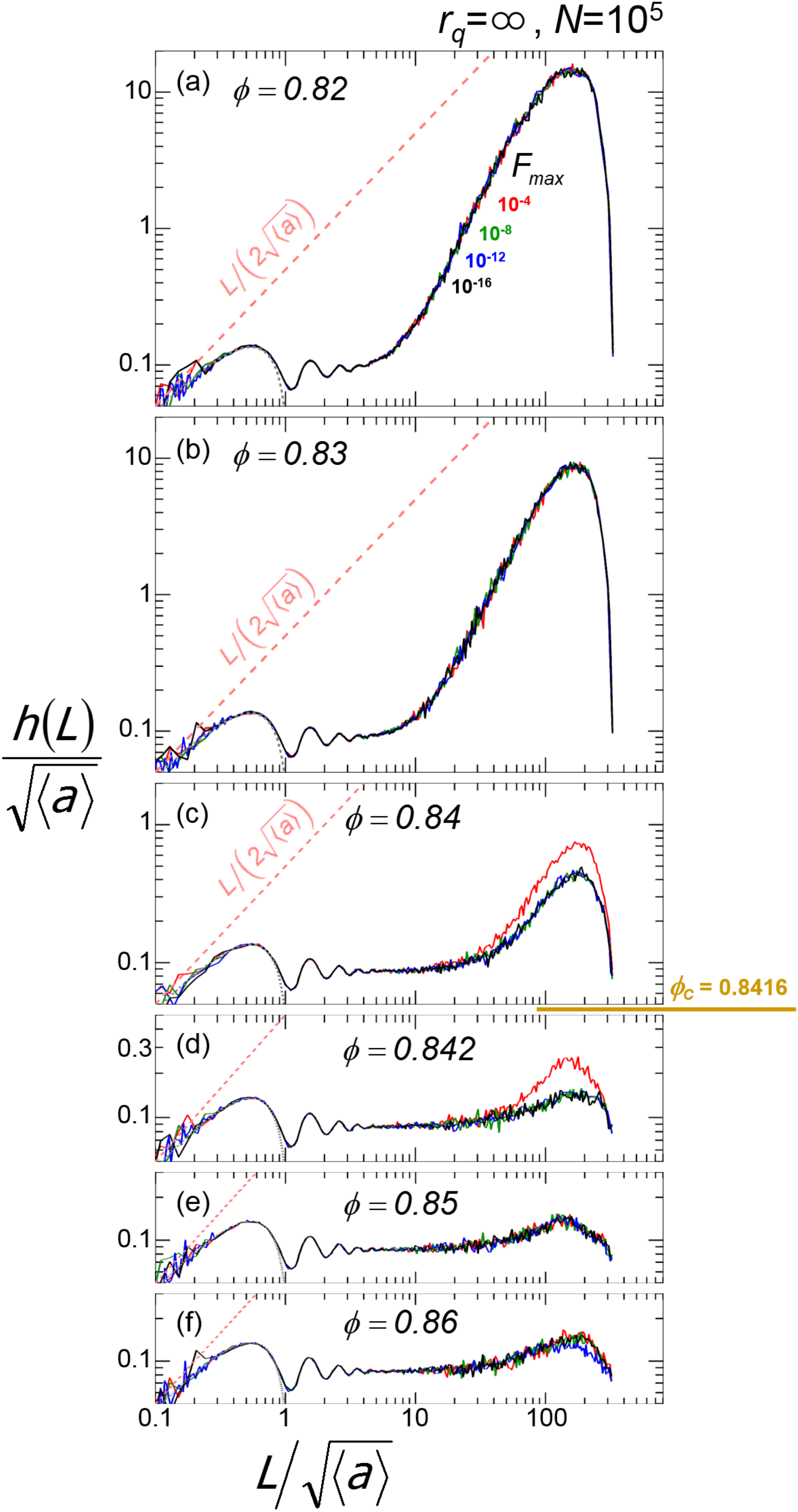}
\caption{The hyperuniformity disorder length versus measuring window side length for several values of the maximum allowed force residue at global packing fraction as labeled. The data for $F_{max} = \{10^{-4}, 10^{-8}, 10^{-12}, 10^{-16}\}$ are plotted in red, green, blue, black respectively and each are an average from 5 configurations of $N=10^5$ particles. The systems are either unjammed [parts (a-c)] or jammed [parts (d-f)] with a critical packing fractions of $\phi_c =0.8416 \pm 0.0003$. Only near the jamming transition does the value of $F_{max}$ affect the data; however, our choice of $F_{max}=10^{-12}$ is safe at all $\phi$.}
\label{F_comp}
\end{figure}

Results for $h(L)$ spectra are plotted in Fig.~\ref{F_comp} for the aforementioned values of $F_{max}$ at various packing fractions below and above the jamming transition at $\phi_c \approx 0.8416$. The data in parts (a),(b) where $\phi \ll \phi_c$ and in parts (e),(f) where $\phi \gg \phi_c$  collapse regardless of $F_{max}$. The data in parts (c),(d) where $\phi \approx \phi_c$ show that configurations made with $F_{max}=10^{-4}$ are less uniform than those made for $F_{max} \leq 10^{-8}$.  However, the value of $L_c$ is nearly unchanged.  Since value of $F_{max}$ used in the main text is many orders of magnitude below the value of $10^{-4}$ that yields improperly equilibrated configurations, we conclude that the large-$L$ Poissonian fluctuations are not an artifact of poor equilibration in the simulations.

%=====================
%=====================

%------------------------------------------
\subsection{Influence of Initial Particle Placements}

\begin{figure}[ht]
\captionsetup{justification=raggedright,singlelinecheck=false,margin=20pt,font=small}
\includegraphics[width=3 in]{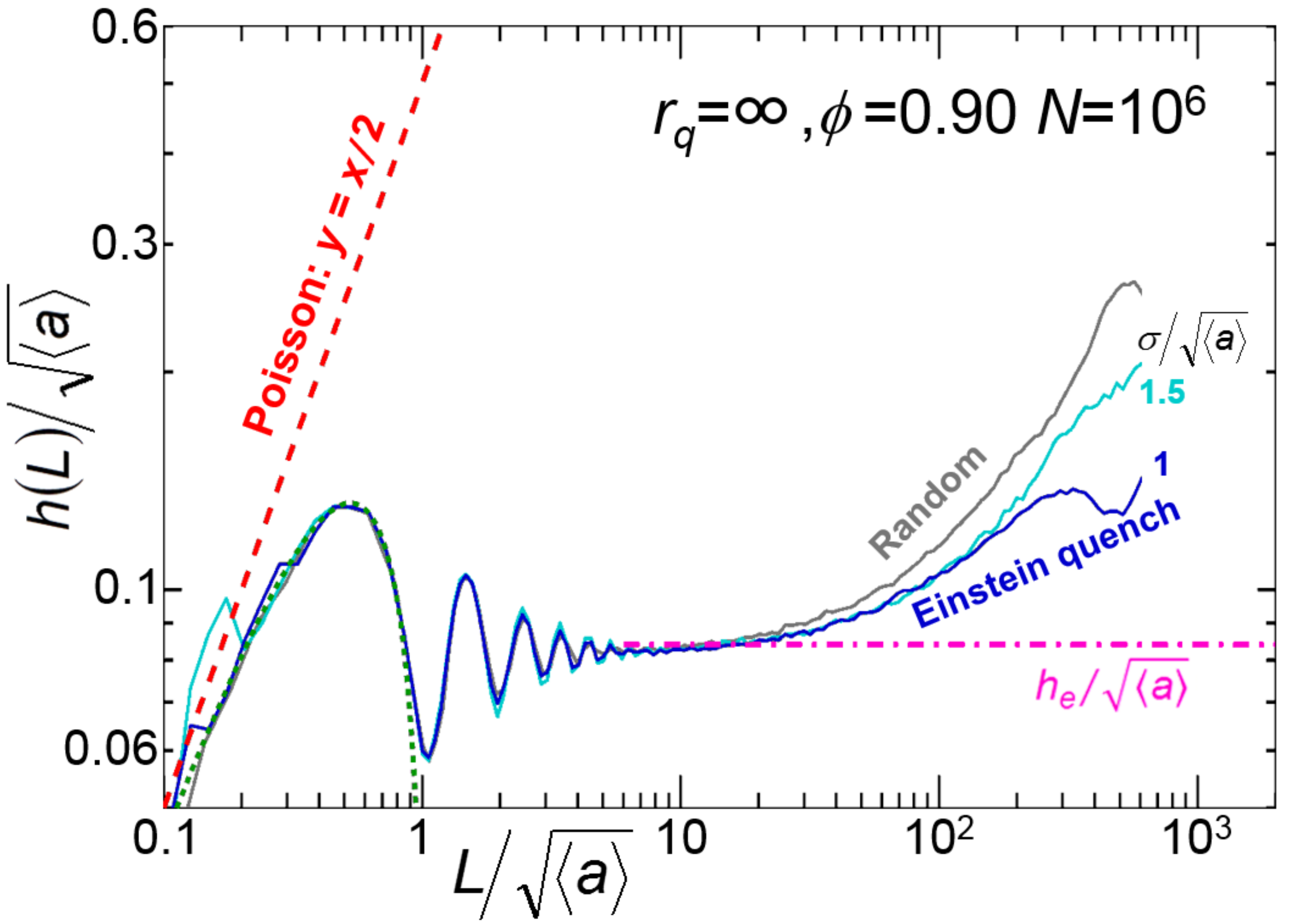}
\caption{ Hyperuniformity disorder length versus measuring window side length for packings quenched from different initial seeding. Data for a standard quench configuration like those described in the main text are plotted in gray and data for Einstein quench configurations are plotted in blue with $\sigma/\sqrt{\left<a\right>}$ as labeled. The Einstein quench curves show that changing the magnitude of the particle displacement does not change the uniformity of the final configuration. Regardless of initial seeding none of the final configurations are hyperuniform but the Einstein quench paterns are slightly more ordered.}
\label{h_seed_dependent}
\end{figure}

For the systems analyzed in the main text, the particles were initially placed into the simulation box with totally random positions.  Here we investigate whether this random seeding into a far-from-equilibrium $h(L)=L/2$ configuration contributes to preventing hyperuniformity at long length scales.  For this, we compare spectra for the equilibrated configurations for three systems with $\{r_q=\infty, \phi=0.9, N=10^6\}$ that result from initializing with random versus Einstein patterns created by randomly kicking the particles from a NaCl lattice with r.m.s.\ displacement of $\sigma$ in each dimension.  As seen in Fig.~\ref{h_seed_dependent}, configurations starting from $\sigma/\sqrt{\langle a\rangle}=\{1, 1.5\}$ have very similar spectra to those starting from a random seeding.  Befitting the initial conditions, the Einstein patterns have slightly larger oscillations at $L=\mathcal O(\sqrt{\langle a\rangle})$ and slightly smaller long-ranged fluctuations.  But certainly, the configurations are still not hyperuniform.  Therefore we conclude that the initial seeding is not responsible for lack of hyperuniformity above jamming.

%=====================
%=====================

%------------------------------------------
\subsection{Influence of Window Shape} \label{WindowShape}

\begin{figure}[ht]
\captionsetup{justification=raggedright,singlelinecheck=false,margin=20pt,font=small}
\includegraphics[width=3in]{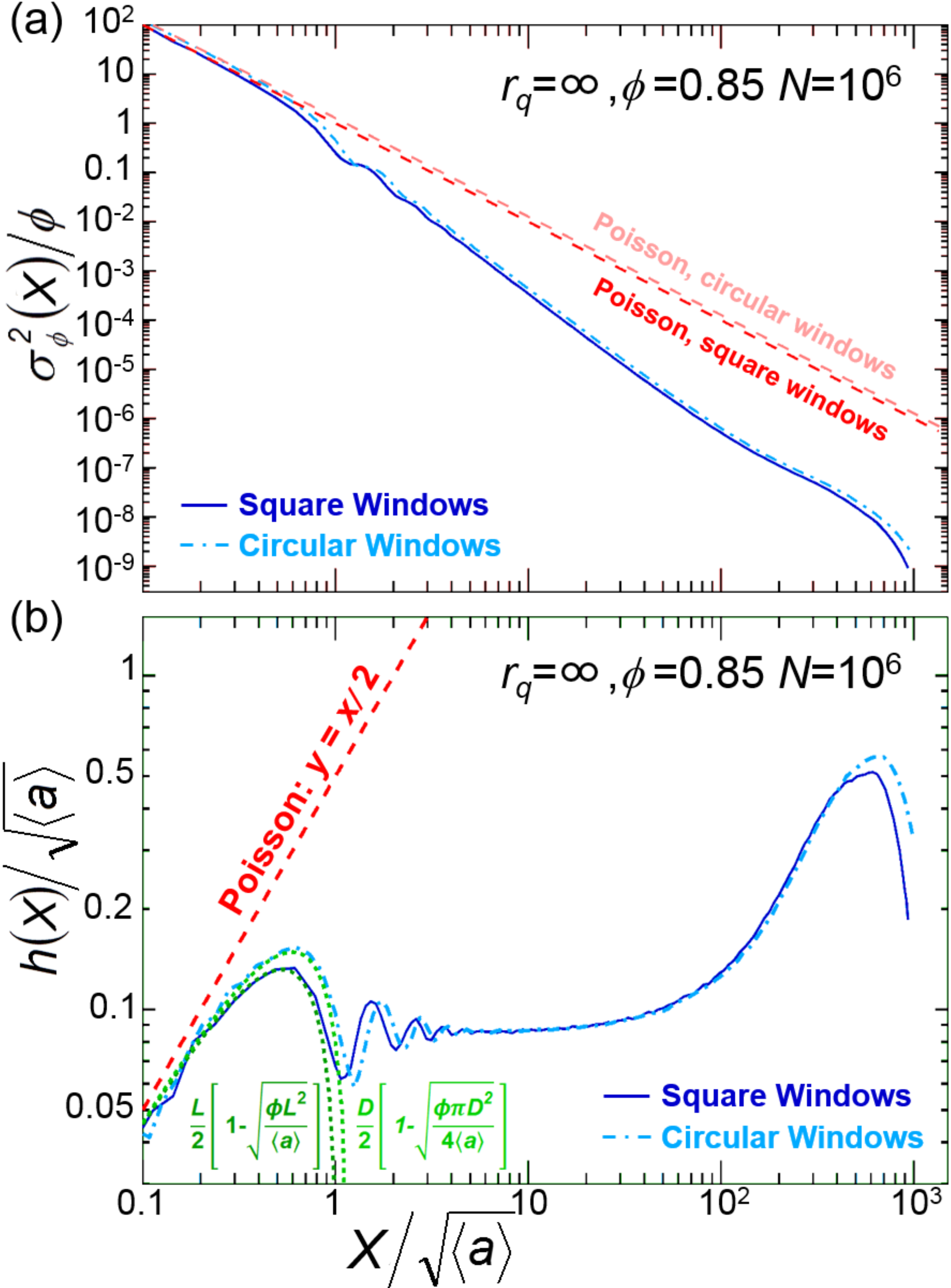}
\caption{ Area fraction variance (a) and hyperuniformity disorder length (b) versus relative measuring window size calculated for either square or circular measuring windows. Data are collected from systems with quench rate, global packing fraction, and number of particles as labeled. $X$ is either the window side length $L$ for square windows or the window diameter $D$ for circular windows. In part (a) the relative variance for circular windows lies slightly above that for the square windows for all $X$ but in part (b) the spectra of hyperuniformity disorder lengths collapse regardless of window shape for large $X$. Expectations for Poisson patterns (red dashed lines) depend on window shape for the relative variance but not for the hyperuniformity disorder length. In both parts the data depend on window shape for small $X$; this is expected because in the separated particle limit (where only 1 or 0 particles land in a window) differences in the ratio between the window area and average particle size matter \cite{ATCpixel}. In part (b) the expectations for the separated particle limit (green dotted curves) for different window shapes match the data exactly until $X$ exceeds the limit.}
\label{h_circ_vs_square}
\end{figure}

As remarked in the main text and in Footnote~\cite{OtherWindows}, we chose hypercubic windows of edge-length $L$ since the surface-to-volume ratio is readily expressed in arbitrary dimension.  This choice is also natural for pixelated data coming from digital video experiments.  However, much previous work uses hyperspherical windows of diameter $D$.  So in Fig.~\ref{h_circ_vs_square} we compare $\sigma_\phi^2(X)$ and $h(X)$ spectra for a large two-dimensional jammed packing, where $X$ is $L$ for square windows and $D$ for circular windows.  The variance data for the different window shapes are slightly offset, due to the window area difference.  By contrast the $h(X)$ data all start at $X/2$,  exhibit slightly different oscillations at $X=\mathcal O(\sqrt{\langle a\rangle})$, and converge to the same plateau and long-range behavior.  We conclude that choice of window shape is not crucial.  However, as seen in Fig.~\ref{h_circ_vs_square}, finite size effects are slightly smaller for circular windows of diameter $D$ than for square windows of width $L=D$ because the former have smaller area.

%=====================
%=====================

%------------------------------------------
\subsection{Influence of System Size} \label{SysSize}

\begin{figure}[ht]
\captionsetup{justification=raggedright,singlelinecheck=false,margin=20pt,font=small}
\includegraphics[width=3 in]{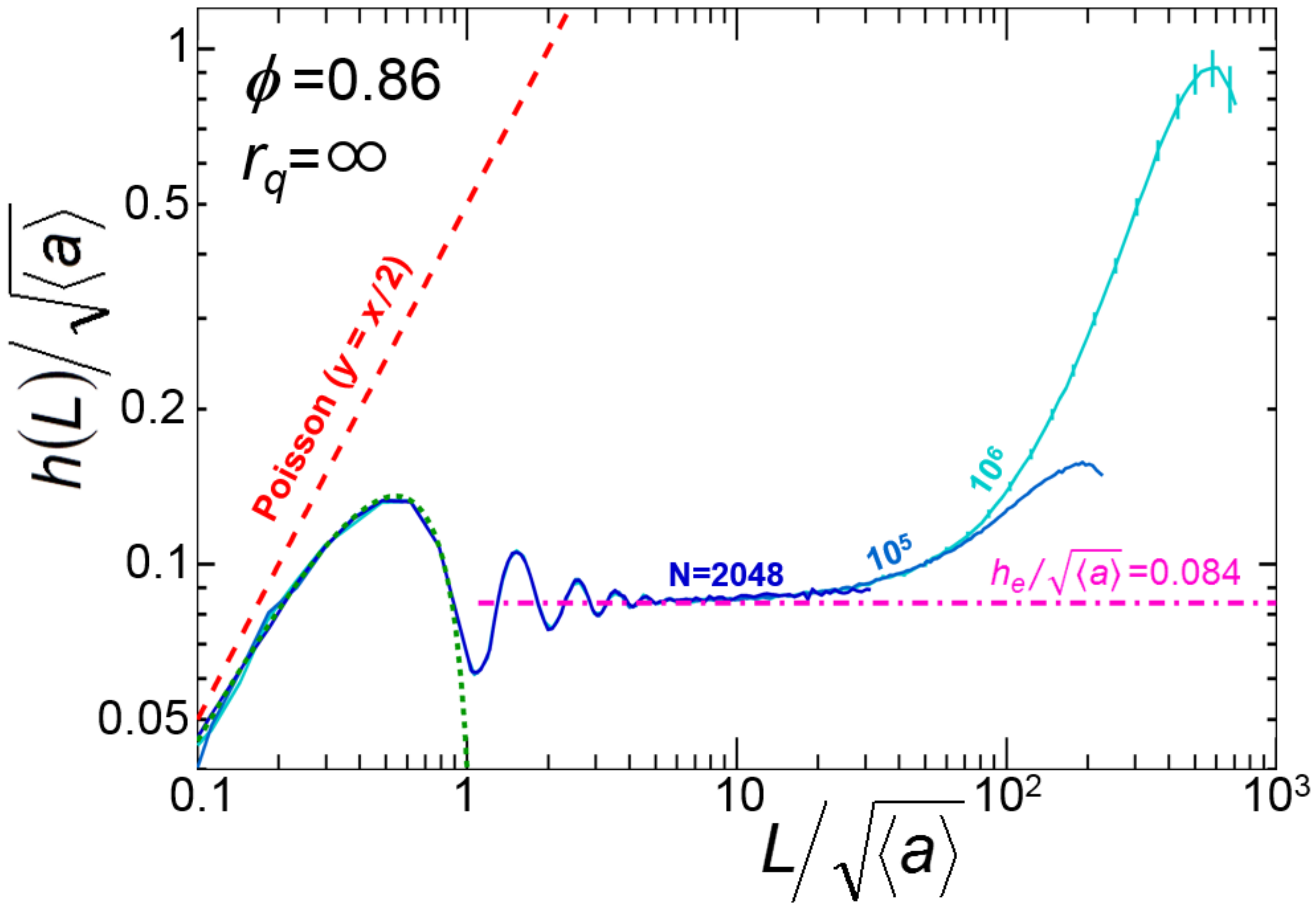}
\caption{ Hyperuniformity disorder length spectra for soft disk packings above jamming by the infinite quench rate protocol.  Data for $N=\{2048, 10^5, 10^6\}$ particles are an average of $\{200, 5, 1\}$ configuration(s), respectively.  These spectra overlap until finite-size artifacts pull $h(L)$ noticeably downwards at $L$ greater than about half the system width. All data are truncated beyond 0.7 times the system width.}
\label{SystemSize}
\end{figure}

Comparing the $h(L)$ spectra for $r_q=\infty$ configurations below jamming in Figs.~\ref{datastacks}h-i it is evident that system size has an effect. Namely, the spectra are not extensive and the larger systems are more random.  However when comparing systems above jamming this ``bigger is different" effect is not seen and the $h(L)$ spectra collapse regardless of system size.  This is demonstrated in Fig.~\ref{SystemSize}, where good collapse is found except where the variance is systematically and erroneously suppressed by finite size effects. 

%=====================
%=====================

%------------------------------------------

\subsection{Influence of Image Size on FFT Output} \label{ImiSize}

The definition we use in the main paper for the spectral density is $\chi(q) = {  \left(  \sum v_j e^{i {\bf q}\cdot{\bf r}_j}~\sum v_k e^{-i {\bf q}\cdot{\bf r}_k}\right) }/{\sum v_j^2 }$ where $q=|{\bf q}|$ for isotropic packings and the sums are over all particles in the system. The numerator is evaluated by Fast Fourier Transform (FFT) on a square digital image consisting of $L_{image} \times L_{image}$ pixels. Our configurations already exist in a square $1 \times 1$ box with periodic boundary conditions, and so to make images we can analyze by FFT we only need to rescale the initial $x$ and $y$ positions and particle radii by the image width $L_{image}$. Then to keep with the central point representation the $(i,j)$ pixel corresponding with the rescaled $(x,y)$ position of a particle is given an intensity equal to the rescaled particle area. This process can be done for any choice of $L_{image}$ but the FFT works best for images with sizes using powers of 2 and so $L_{image}=\{2^{12},2^{13},2^{14}\}$ is used for configurations with $N=\{2048,10^5,10^6\}$, respectively.

While the rescaling process is simple it may lead to significant and erroneous particle overlap if the image is too small. This is because the central point representation allows the particle areas to be rescaled to arbitrary precision but the $(x,y)$ positions have to be rounded to their nearest integer due to space discretization. If the image is small enough that the rounded particle positions are closer together than the rescaled particle radii then false random overlaps are introduced. These overlaps would corrupt our analysis of positional order so it is critical that we use images large enough that increasing the image size does not appreciably change the FFT output.

\begin{figure}[ht]
\captionsetup{justification=raggedright,singlelinecheck=false,margin=20pt,font=small}
\includegraphics[width=3in]{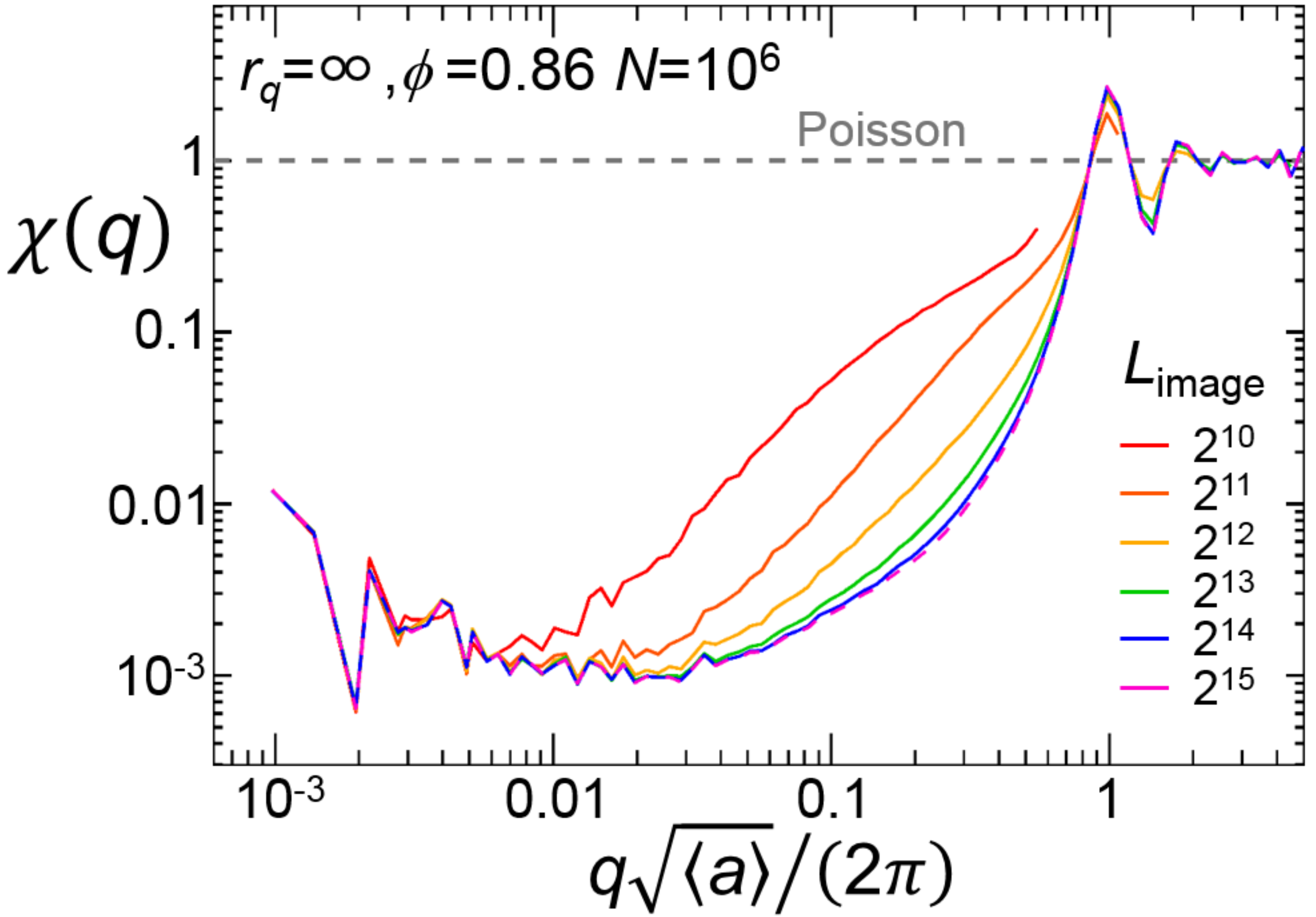}
\caption{Spectral density versus wavenumber, obtained by FFT of $L_{image}\times L_{image}$ pixelated image representations of particles positions for a configuration generated by the infinite quench protocol for $N=10^6$ particles at $\phi=0.86$. The curves for $L_{image}=\{2^{14},2^{15}\}$ are nearly identical so the latter curve is dashed to make both curves distinguishable. The dashed gray line is the expectation for a completely random system. }
\label{spec_sys_size}
\end{figure}

In Fig.~\ref{spec_sys_size} we plot the dimensionless spectral density $\chi(q)$ versus dimensionless wavenumber $x=q\sqrt{\langle a\rangle}/(2\pi)$ for one configuration made by the infinite quench protocol with $N=10^6$ particles and $\phi=0.86$, analyzed at many image sizes.   One immediately notices that the range of $x$-values is affected by the system size for large $x$ only. This is because both $q$ and $\left< a \right>$ change with system size:  Possible wavenumbers are $q = 2 \pi n / L_{image}$ where $n=1,2 ... N_L/2$ and $N_L$ is the number of pixels along one side of the image; note the max is $N_L/2$ because we radially average from the center of the FFT output image to find $\chi \left( q \right)$. The value of $\left< a \right>$ depends on image size and system specifications:  $\left< a \right>= 2\phi {L_{image}}^2 / \left[ N_{tot} \left( 1 + c \right) \right]$ where $\phi$ is the packing fraction, $N_{tot}$ is the total number of particles and $c$ is the large to small particle size ratio. Normalizing $q_{min}$ and $q_{max}$ by $\sqrt{\left< a \right>}/ (2 \pi)$ gives the bounds $x_{min}=\sqrt{2\phi/ \left[ N_{tot} \left( 1 + c \right) \right]}$ and $x_{max}= \left( N_L / 2 \right) \sqrt{2\phi/ \left[ N_{tot} \left( 1 + c \right) \right]}$. The minimum $x$-value is a constant regardless of image size but the maximum endpoint changes because it depends on $N_L$.

As for the spectral density neither the small-$x$ ($x \leq 0.01$) nor the large-$x$ ($x \geq 1$) behavior are affected by the image size for $L_{image}>2^{12}$. The large-$x$ behavior shows oscillations around $\chi \left(q \right)=1$ and the difference in extrema grows for larger images. The small-$x$ behavior shows the data leveling off or possibly increasing slightly; again there are differences in the actual value of $\chi \left( q \right)$ for different $L_{image}$ but as the image size increases so does the number of $x$-values where $\chi \left( q \right)$ are the same. This observation is especially important because the most significant differences between data sets occur at intermediate $q$-values. However for $L_{image} \geq 2^{12}$ the differences are restriced to the magnitude of $\chi \left( q \right)$ and similar behavior, which shows a decaying value of $\chi \left( q \right)$ until about $x=0.03$ where the data level off, is captured. For our analysis we compute $\chi \left( q \right)$ with images that have $L_{image}=2^{14}$ for the $N=10^6$ particle packings and for this image size and greater the data are nearly identical for all $x$. We safely conclude that our spectral density data are not corrupted by image-size effects.

%=====================
%=====================

% Create the reference section using BibTeX:
\bibliography{../HyperRefs}

\begin{thebibliography}{36}
\bibitem [{Oth()}]{OtherWindows}%
  \BibitemOpen
  \href@noop {} {}\bibinfo {note} {We use hypercubic windows so that Eqs.~(1,2)
  can be written simply in any dimension. For step-function windows of volume
  $V_\Omega$ and surface area $A_\Omega$, of any shape and dimensionality,
  Eq.~(2) is more generally $\sigma_\phi^2/\phi \approx \langle v \rangle
  A_\Omega h/ V_\Omega^2$ for $V_\Omega/A_\Omega \gg h$. For $d=2$ circular
  windows, Eq.~(1) becomes $\sigma_\phi^2(R)/\phi = [\langle a\rangle/(\pi
  R^2)] [1 - (1-h/R)^2 ]$ and reduces to $2\langle a\rangle h/(\pi R^3)$ for
  $R\gg h$. For $d=3$ spherical windows, Eq.~(1) becomes $\sigma_\phi^2(R)/\phi
  = [\langle v\rangle/(4\pi R^3/3)] [ 1 - (1 - h/R)^3 ]$ and reduces to
  $\sigma_\phi^2(R)/\phi \approx 9\langle v\rangle h/(4\pi R^4)$ for $R\gg h$.
  See Ref.~\cite{DJDhudls} for non step-function windows, such as a
  Gaussian.}\BibitemShut {Stop}%
\bibitem [{exp()}]{experiment}%
  \BibitemOpen
  \href@noop {} {}\bibinfo {note} {For laboratory data there are two other
  sources of error. First, $h$ is inflated by about half the uncertainty in
  particle positions~\cite{ATCpixel}. Second, uncertainty or unaccounted-for
  polydispersity $\delta v_i$ in species $i$ inflates the measured variance to
  $\sigma_\phi^2(L)/\phi = [\sigma_\phi^2(L)/\phi ]_{\rm actual} +\langle
  \delta v^2 / v\rangle/ V_\Omega$ where $\langle \cdots \rangle$ is the
  $\phi_i$-weighted average of $\delta v_i^2/v_i$. This error is Poissonian and
  inevitably masks true hyperuniformity at large enough windows.}\BibitemShut
  {Stop}%
\end{thebibliography}
\end{document}